\documentclass[english]{article}
\usepackage[T1]{fontenc}
\usepackage[latin9]{inputenc}
\usepackage{geometry}
\usepackage{textcomp}
\usepackage{mathrsfs}
\usepackage{amsmath}
    \newcommand\numberthis{\addtocounter{equation}{1}\tag{\theequation}}
\usepackage{amssymb}
\usepackage{graphicx}
\usepackage{setspace}
\usepackage{nccmath}
\makeatletter

\providecommand{\tabularnewline}{\\}

\newenvironment{lyxlist}[1]
	{\begin{list}{}
		{\settowidth{\labelwidth}{#1}
		 \setlength{\leftmargin}{\labelwidth}
		 \addtolength{\leftmargin}{\labelsep}
		 }}
	{\end{list}}

\usepackage{lmodern}
\usepackage{graphicx}
\usepackage{color}
\usepackage{bbm}
\usepackage{pdflscape}
\usepackage{setspace}
\usepackage{booktabs}
\usepackage{multirow}
\usepackage{amsthm}
\usepackage{hyperref}

\newtheorem{theorem}{Theorem}
\newtheorem{assumption}{Assumption}
\newtheorem{corollary}{Corollary}
\newtheorem{lemma}{Lemma}[section]

\newenvironment{Proof}{%
    \noindent {\scshape{Proof}} \newline \indent
}{%
    \newline
    $\square$
}

\usepackage[round, authoryear]{natbib}
\usepackage{appendix}

\makeatother

\usepackage{babel}
\begin{document}
\title{Spatial Econometrics for Misaligned Data}
\author{Guillaume Allaire Pouliot\thanks{Harris School of Public Policy, University of Chicago. 1307 E 60th
St, Chicago, IL 60637. tel: (773) 702-8400. email: guillaumepouliot@uchicago.edu. website: https://sites.google.com/site/guillaumeallairepouliot/research.
 The R package SpReg implementing the methods is available on the author's website.}}
\maketitle
\begin{abstract}
We produce methodology for regression analysis when the geographic
locations of the independent and dependent variables do not coincide,
in which case we speak of misaligned data. We develop and investigate
two complementary methods for regression analysis with misaligned
data that circumvent the need to estimate or specify the covariance
of the regression errors. We carry out a detailed reanalysis of \citet{maccini2009under} and find economically significant quantitative
differences but sustain most qualitative conclusions.
\end{abstract}
Spatial data analysis has become increasingly popular in the social
sciences. In many applications, data sets providing the specific location
of households, firms, villages, or other economic units are matched
by location to data sets with geographic features such as rainfall, temperature,
soil quality, ruggedness, or air pollution in order to analyze the
impact of such environmental variables on economic outcomes.\footnote{\citet{maccini2009under}, \citet{miguel2004economic}, and \citet{shah2013drought} study the impact of rainfall. \citet{dell2014we} survey applications
using weather data. \citet{fabregas2017institutions} use measurements of soil
nutrients in some locations to make fertilizer recommendations in
others. \citet{nunn2012ruggedness} use terrain ruggedness for identification,
and \citet{chay2003impact} study the impact of air pollution. } Such data underpins important economic research, including policy
responses to droughts, smog outbreaks, poor harvests, and other events.
A typical issue is that the matched data sets will be \emph{misaligned}.
That is, the respective geographical locations of the observations
in the matched data sets do not generally coincide. For instance,
a researcher might observe crop outputs from a  sample of farms in
a large area as well as measurements of rainfall collected over the
same area from several weather stations. The locations of the weather
stations and the farms will generally not coincide, resulting in
misaligned data sets.

\begin{figure}
\begin{center}\includegraphics[scale=0.3]{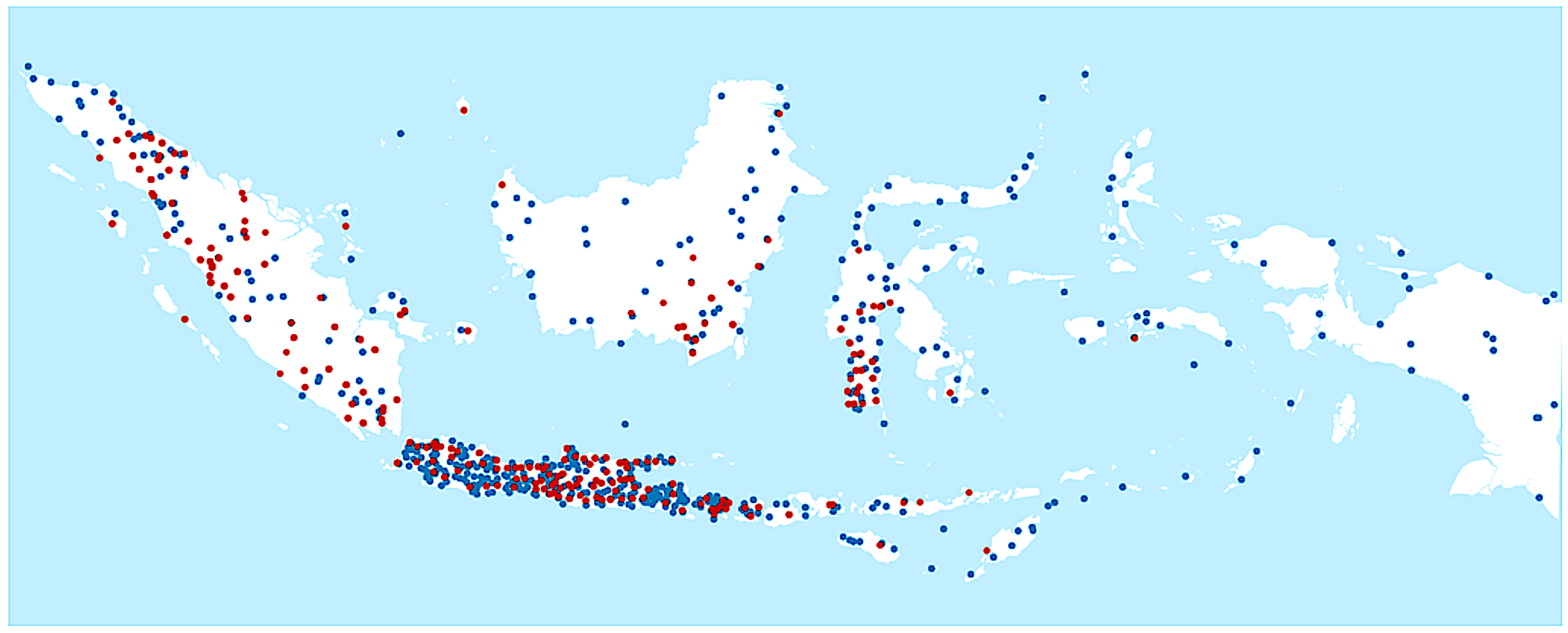}\end{center}

\caption{Map of Indonesia. Geographic locations of rainfall measurements (blue)
and survey data (red) merged in the \citet{maccini2009under} analysis. }
\label{fig:map}
\end{figure}



The approaches commonly used in social sciences to address the misalignment
problem yield inefficient estimates and incorrect confidence intervals,
and are consistent only under implausible assumptions. Popular approaches
for analyzing such data sets involve imputing a level of the regressor for each misaligned observation of the outcome variable
and then proceeding with standard regression analysis. It is common
to impute using either the value of the nearest location,
 the value of the nearest location instrumented with
more distant locations \citep{maccini2009under}, or a distance-weighted
average of nearby locations \citep{shah2013drought}. These methods
impute the regressors in an initial step before considering the outcome
data. Hence we refer to them as ``two-step'' or ``plug-in'' methods. As detailed below and discussed in the simulations and application, two-step methods used in the literature
are often inconsistent and are generally inefficient because they
do not use all the relevant information.

That being said, efficient estimation is challenging in social science
applications because the covariance structure of the regression errors
will typically be unknown, which raises a fundamental identification
issue since, for instance, the Gaussian maximum likelihood estimator
requires stipulating the covariance of the regression errors for point
estimation \citep{pouliot2016missing}.

Specification of the covariance model for the regression errors is
arguably the most onerous assumption the researcher needs to make.
We may worry that the regression error covariance does not have the
nice structure common to geophysical variables, e.g., since we ineluctably
omit variables in the regression function. Consider, for example,
pseudo-replicates (also referred to as random effects or clustered
errors): it may be that some regions and their crops are affected
by toxic emissions, but we do not know which regions. Our inability
to identify which observations share this regional shock prevents
us from identifying the correct covariance model. This motivates an
estimation scheme relieved of that specification burden.

Indeed, the researcher may or may not have a suitable model for the
(spatial) covariance of regression errors. We thus require methodology
that will dispense with the requirement to model the regression error
covariance structure while preserving as much statistical power as
possible.

We propose an easy-to-implement two-step method, based on a Krig-and-regress
approach, that does not require stipulation of the regression error
covariance and produces standard errors that account for uncertainty
due to first stage estimation.

We further argue that a suitable choice of moments, excluding the
regression error covariance, identifies the regression coefficients,
and we construct a minimum-distance estimator based on these moments.
Jointly estimating all coefficients, this estimator brings about efficiency
gains.

These two methods should be thought of as complementary. As further
detailed below, while the two-step Krig-and-regress estimator is substantially
simpler and more transparent, efficiency gains may obtain from the
minimum-distance estimator, especially on gridded data.

Reproducing and extending the cross-validation exercise of \citet{madsen2008regression}, we find that both estimators are competitive, but the
more efficient minimum-distance estimator outperforms the Krig-and-regress
estimator when the locations are regular, i.e., a subset of a lattice. 

We reanalyze the influential data set of \citet{maccini2009under} and
find that the analysis benefits from the use of our methods, as those
yield some statistically and economically significant changes in the
value of key parameter estimates.

\subsection*{Problem Set-Up}

We are interested in the regression coefficient $\beta$ in the spatial
regression problem

\begin{equation}
\mathbf{Y}=\mathbf{R}_{\mathrm{true}}\beta+F\gamma+\epsilon,
\end{equation}
where $\mathbf{Y}=Y(x)$ is an $N$-tuple $(Y(x_1), \dots, Y(x_N))^T$ and 
\begin{equation}
\mathbf{R}_{\mathrm{true}}=R(x)=(R(x_{1}), \dots ,R(x_{N}))^{T}
\end{equation}
is drawn from a stationary 
random field\footnote{We further speak of  $\left\{ R(x):x\in\mathbb{R}^{d}\right\} $ as a Gaussian
random field if, for any choice of vector of locations $(x_{1},...,x_{n})$,
the random vector $(R(x_{1}),...,R(x_{n}))$ is distributed multivariate
normal. The practical usefulness of Gaussian random fields to model
rainfall data has long been established, see in particular \citet{phillips1992comparison} and \citet{tabios1985comparative}.  Nowhere will we, however, require the correct specification of Gaussianity.} $R(\cdot)$ evaluated at geographic locations $x_{i}\in\mathscr{\mathscr{D}}\subset\mathbb{R}^{2}$,
$i=1,...,N$.
The error vector $\epsilon \in \mathbb{R}^{N}$ has mean zero and unknown covariance matrix $\Sigma$ and is independent of $R$.
The outcome vector $\mathbf{Y}$ and the matrix of controls $F$ are
observed, and the vector of regression coefficients of the controls $\gamma$ is unknown.  The matrix of control
variables $F$ may include covariates such as age, location dummies, etc.  The difficulty is that $\mathbf{R}_{\mathrm{true}}$ is not observed. However, the $M$-tuple
\begin{equation}
\mathbf{R}^{*}=R(x^{*})=(R(x_{1}^{*}), \dots, R(x_{M}^{*}))^{T},
\end{equation}
 with $x_{i}^{*}\in\mathscr{\mathscr{D}}\subset\mathbb{R}^{2}$, $i=1,...,M$,
is observed. That is, although the outcome variable data $Y(x)$ (e.g.,
crop yields at farm locations) is not sampled at the same locations
as the independent variable data $R(x^{*})$ (e.g., rain measured
at fixed weather stations), it is $R$ evaluated at the same locations
as that of the outcome variable, that is $R(x)$, which enters the
regression function.

The random field $R(\cdot)$ has mean function $m:\mathscr{\mathscr{D}}\rightarrow\mathbb{R}$
and covariance function $K_{\theta}:\mathscr{\mathscr{D}}\times\mathscr{\mathscr{D}}\rightarrow\mathbb{R}_{+}$,
where $\theta$ indexes a parametric model for the covariance function. The mean and variance of the observables are then

\begin{equation}
E\left(\begin{array}{c}
\mathbf{Y}\\
\mathbf{R}^{*}
\end{array}\right)=\left(\begin{array}{c}
m(x)\beta+F\gamma\\
m(x^{*})
\end{array}\right)
\end{equation}
and
\begin{equation}
V\left(\begin{array}{c}
\mathbf{Y}\\
\mathbf{R}^{*}
\end{array}\right)=\left(\begin{array}{cc}
\mbox{\ensuremath{\beta}}^{2}\mathbf{K}+\Sigma & \beta\mathbf{\bar{K}}\\
\beta\mathbf{\bar{K}}^{T} & \mathbf{K^{*}}
\end{array}\right),
\label{eq:varianceobservable}
\end{equation}
respectively, where $\mathbf{K}=K_{\theta}(x,x)=V_{\theta}(R(x))\in \mathbb{R}^{N 
\times N}$,
$\mathbf{\bar{K}}=K_{\theta}(x,x^{*})=Cov_{\theta}(R(x),R(x^{*}))\in \mathbb{R}^{N 
\times M}$
and $\mathbf{K^{*}}=K_{\theta}(x^{*},x^{*})=V_{\theta}(R(x^{*}))\in \mathbb{R}^{M 
\times M}$
for some $\theta\in\Theta$.

In the absence of rainfall measurements at the locations of outcomes,
the identifying assumption is that we have a parametrized covariance
function, and know $Cov_{\theta}(\mathbf{R}_{\mathrm{true}},\mathbf{R}^{*})$
up to the value of a small-dimensional parameter vector $\theta$,
which we estimate consistently. This allows, for instance, the construction
of a best linear unbiased predictor for unobserved rainfall.

For our purposes, it will generally be the case that $m$ is constant with respect to the location $x$,
and thus the mean parameter of the random field $R$ can be absorbed
in the constant vector (for the intercept) in the matrix of control
variables $F$. Hence, we are concerned throughout with the mean and variance

\begin{equation}
E\left(\begin{array}{c}
\mathbf{Y}\\
\mathbf{R}^{*}
\end{array}\right)=\left(\begin{array}{c}
F\gamma\\
\mathbf{m^{*}}
\end{array}\right)
\end{equation}
and
\begin{equation}
V\left(\begin{array}{c}
\mathbf{Y}\\
\mathbf{R}^{*}
\end{array}\right)=\left(\begin{array}{cc}
\mbox{\ensuremath{\beta}}^{2}\mathbf{K}+\Sigma & \beta\mathbf{\bar{K}}\\
\beta\mathbf{\bar{K}}^{T} & \mathbf{K^{*}}
\end{array}\right),
\label{eq:variance}
\end{equation}
where the coefficient of interest, $\beta$, only appears in the covariance.

\subsection*{Related Literature}

This article pertains to different segments of the literature. First
and foremost, we provide methods for applied researchers. As detailed above, even careful applied work \citep{ maccini2009under, shah2013drought} relies on \emph{ad hoc }approaches because econometric
methodology has not caught up to the needs of applied economists.
This paper intends to address those methodological needs. It speaks
to a well-established literature on generated regressors \citep{pagan1984econometric, murphy1985estimation}, and looks at problems in which the regressors to impute are amenable to imputation using the best
linear predictor, e.g., their law well approximated by a Gaussian random
field. Gaussian random fields and best linear prediction, also known as Kriging,
are well-studied in geostatistics \citep{gelfand2010handbook, stein2012interpolation}
where, however, interest is concentrated on interpolation. This paper
relates the two literatures and leverages geostatistical methodology
and results to provide robust and accurate methods for economists
carrying out regression analysis with misaligned data.

The central limit theorem we obtain for inference with the minimum-distance
estimator builds on results in spatial statistics. \citet{lahiri2002asymptotic} give a central limit theorem for empirical variogram estimators
when the data is on a lattice. We leverage results from \citet{lahiri2003central},
who gives a family of central limit theorems for spatial statistics,
in order to extend the asymptotic theory for the empirical variogram
estimators to the case of irregularly spaced data -- which
is increasingly common in development economics as well as other fields
of applied economics.

An alternative asymptotic theory is laid out in the work of \citet{jenish2009central, jenish2012spatial}. Their theory is very general, even allowing
for nonstationarity.  Definitions of
``spatial mixing'' and rate assumptions are not critically different.
As suggested by the application to minimum-distance estimation on
lattice data \citep{lahiri2002asymptotic}, the general results in \citet{lahiri2003central}
are immediately relevant to the analysis of our minimum-distance estimator.
That being said, we have found, for instance, that the approach of \citet{jenish2012spatial} is more convenient for developing maximum likelihood
asymptotic distribution theory, as exemplified in the quasi-likelihood
case by \citet{qu2017qml} and the well-specified case by \citet{xu2015maximum}.

\subsection*{Outline}

The remainder of the article is divided as follows. Section \ref{sec:background} presents
and discusses key concepts for the analysis. Section \ref{sec:method} presents Krig-and-regress
and minimum-distance methods dealing with misaligned data.
Section \ref{sec:madsen} studies the comparative performance of the considered estimators
in the cross-validation exercise of \citet{madsen2008regression}. Section
\ref{sec:maccini} reanalyzes the misaligned data set in \citet{maccini2009under}. Section
\ref{sec:conclusion} discusses and concludes. Technical material is deferred to the Appendix.

\section{Key Concepts and Background Material}\label{sec:background}

Two-step methods for regression analysis of misaligned data consist
in first predicting the misaligned covariates at the outcome locations
where they are not observed, thereby generating an aligned data set,
and then proceeding to spatial regression analysis with this
generated data set. The first step, which consists of predicting the
missing independent variables, requires the choice of an interpolation
method. Nonparametric methods, such as approximation by the average
of a given number of nearest neighbors, may be used. However, when
the misaligned variable can be modeled as following, or approximately
following, the law of a Gaussian random field, Kriging generally affords
the researcher more accurate interpolation \citep{gelfand2010handbook}.

Kriging (named after the South African mining engineer D. G. Krige)
consists in using the estimated best linear unbiased predictor for
interpolation. It can be developed as follows \citep{stein2012interpolation}. The random
field of interest, $R$, is assumed to follow the model 
\[
R(x)=s(x)^{T}\rho+\varepsilon(x),
\]
$x\in\mathscr{D}\subset\mathbb{R}^{2}$, where $\varepsilon$ is a
mean zero random field, $s$ is a known function with values in $\mathbb{R}^{p}$
and $\rho$ is a vector of $p$ coefficients. We observe $R$ at
locations $x_{1}^{*},x_{2}^{*},...,x_{M}^{*}$. That is, we observe
$\mathbf{R}^{*}=\left(R(x_{1}^{*}),R(x_{2}^{*}),...,R(x_{M}^{*})\right)$
and need to predict $R(x_{0})$. With $\rho$ known, the best linear
predictor (BLP) is
\[
s(x_{0})^{T}\rho+\mathbf{k}^{T}\mathbf{K}^{-1}(\mathbf{R}^{*}-\mathbf{S}\rho),
\]
where $\mathbf{k}=Cov(\mathbf{R}^{*},R(x_{0}))$, $\mathbf{K}=Cov(\mathbf{R}^{*},\mathbf{R}^{*T})$
and $\mathbf{S}=(s(x_{1}^{*}),s(x_{2}^{*}),...,s(x_{M}^{*}))^{T}$.
Of course, the mean parameter $\rho$ is, in general, unknown. If
$\rho$ is replaced by its generalized least-squares estimator,
$\hat{\rho}=\left(\mathbf{S}^{T}\mathbf{K}^{-1}\mathbf{S}\right)^{-1}\mathbf{S}^{T}\mathbf{K^{-1}}\mathbf{R}^{*}$
(under the assumption that $\mathbf{K}$ and $\mathbf{S}$ are of
full rank), we obtain the best linear unbiased predictor (BLUP) for
$R(x_{0})$. Again, in general, the covariance structure will be unknown,
and $\mathbf{k}$ and $\mathbf{K}$ will be replaced by estimates
$\hat{\mathbf{k}}$ and $\hat{\mathbf{K}}$. The resulting plug-in
estimator will be called the estimated BLUP (EBLUP). Prediction with
the BLUP and EBLUP are both referred to as Kriging. As far as this
article is concerned, the covariance structures will always be \emph{a
priori} unknown, and Kriging will refer to prediction with the EBLUP. 

There are many choices for the covariance functions \citep{gelfand2010handbook}, and we present three of them in the isotropic case in which
only the distance $d$ between the covariates determines their covariance:
the exponential covariance function
\begin{equation}
K_{\mathrm{exp}}(d)=\theta_{1}\exp\left(-d/\theta_{2}\right),
\end{equation}
the Gaussian covariance function
\begin{equation}
K_{\mathrm{Gaussian}}(d)=\theta_{1}\exp\left(-d^{2}/\theta_{2}^{2}\right),
\label{eq:gaussiancov}
\end{equation}
and the Mat\'ern covariance function

\begin{equation}
K_{\mathrm{Mat\acute{e}rn}}(d)=\theta_{1}\frac{\left(d/\theta_{2}\right)^{\nu}\mathcal{K}_{\nu}\left(d/\theta_{2}\right)}{2^{\nu-1}\Gamma(\nu)},
\end{equation}
where $\mathcal{K}_{\nu}$ is the modified Bessel function of the
second kind of order $\nu$ \citep[sec. 9.6]{abramowitz1964handbook}.
All functions have positive parameters $\theta_{1}$ and $\theta_{2}$,
which are the sill and range, respectively. The sill parameter should
be thought of as controlling the scale of the covariance, and the
range should be thought of as controlling how fast the covariance
decays over distance. The Mat\'ern function has an additional parameter
$\nu$, which controls smoothness.

\citet{jiang1997derivation} and \citet{stein2012interpolation} present an alternative derivation of
the BLUP as the best predictor, under normality, based on all error
contrasts. An excellent theoretical treatment of the topic can be
found in \citet{stein2012interpolation}. \citet{cressie2015statistics} and \citet{diggle2007springer}
offer a more applied treatment of the topic. \citet{matheron1962traite} is a
classic reference.

Kriging, or the best linear prediction of missing variables, naturally
extends regression, or the estimation of best linear prediction of
the outcome variable, to the misaligned case. Indeed, we want to estimate
the best linear predictor\footnote{The best linear predictor is defined analogously to the conditional
expectation. We may define the conditional expectation as the best
predictor $f(R^{*})$ minimizing $E\left[\left(R-f(\mathbf{R}^{*})\right)^{2}\right]$.
Likewise, the best linear predictor is the linear function $\alpha^{T}\mathbf{R}^{*}$
minimizing $E\left[\left(R-\alpha^{T}\mathbf{R}^{*}\right)^{2}\right]$
over all $\alpha\in\mathbb{R}^{M}$. } $E^{*}\left[Y\left|R\right.\right]=\beta_{0}+\beta R$ (omitting
all other control variables for simplicity of exposition) but only
observe $Y$ and $R^{*}$. However, note that 
\begin{eqnarray*}
E^{*}\left[Y\left|R^{*}\right.\right] & = & E^{*}\left[E^{*}\left[Y\left|R\right.\right]\left|R^{*}\right.\right]\\
 & = & \beta_{0}+\beta E^{*}\left[R\left|R^{*}\right.\right],
\end{eqnarray*}
hence a consistent estimate of $E^{*}\left[R\left|R^{*}\right.\right]$
will deliver a consistent estimate of the best linear predictor $E^{*}\left[Y\left|R^{*}\right.\right]$ and,
in particular, of $\beta$.

To be sure, although two-step methods commonly used in the literature
(e.g. nearest neighbors, two-stage least squares) are not in general
consistent, Krig-and-regress is a consistent two-step method. However,
Krig-and-regress does not make efficient use of the data; in the first
stage, only the covariance $Cov\left(\mathbf{R}^{*},\mathbf{R}^{*}\right)$
is used to estimate the rainfall covariance $\theta$ and build the
EBLUP, but $\theta$ is also informed by the second stage statistic
$Cov\left(\mathbf{Y},\mathbf{R}^{*}\right)=\beta\mathbf{\bar{K}}$.
Minimum-distance estimation offers a natural alternative and uses
both moments to jointly estimate $\theta$ and $\beta$.

\section{Methodology}\label{sec:method}

One of our main motivations is to build principled methods for misaligned
regression that are pivotal with respect to $\Sigma$, the covariance
of the regression errors. If one wants to circumvent altogether the
need to model the covariance of the regression errors, the two-step
bootstrap of Subsection \ref{ssec:krigregress} delivers consistent estimation and valid
inference. If one has a model for the covariance of the regression
errors but does not want to rely on it for point estimation, the minimum-distance
estimator may be employed. With the latter estimator, covariance parameters
appearing in both the first and second stage problems are estimated
only once, jointly on the first and second stage data, and specification
of the covariance of regression errors, $\Sigma$, is only required
for inference. If one wants to avoid specifying $\Sigma$ even for
inference with the minimum-distance estimator, a heuristic approximate
Bayesian computation (ABC) approach to inference is provided which
does not require stipulating $\Sigma$. Table \ref{tab:method} summarizes when each
method should be preconized. The R package SpReg implementing both methods is available on the author's webpage.\footnote{https://sites.google.com/site/guillaumeallairepouliot/research}

\begin{table}

\centering
\begin{tabular}{cccc}
\toprule 
 & Need $\Sigma$ for estimation & Need $\Sigma$ for inference & Joint estimation\tabularnewline
\cmidrule{2-4}
Minimum Distance   & no & yes & yes\\
Minimum Distance (ABC) & no & no & yes\\
Two-Step Bootstrap & no & no & no \\
\bottomrule 
\end{tabular}

\caption{Recommended Methods. Columns, from left to right, indicate if a model
for the covariance of regression errors must be stipulated for point
estimation, for inference, and whether all parameters are jointly
estimated on the entire data set.}
\label{tab:method}
\end{table}

\subsection{Krig-and-Regress}\label{ssec:krigregress}

The main motivation for resorting to a two-step method such as Krig-and-regress
is the desire to avoid specifying a model for the covariance of the
regression errors, $\Sigma$. Indeed, the researcher may conclude
that modeling the covariance of the regression errors is too restrictive.
For instance, specification of different but equally credible covariance
structures may yield tangibly different maximum likelihood estimators
\citep{pouliot2016missing}. Furthermore, researchers may preconize the two-step
method for its simplicity and ease of implementation.

Explicitly, the Krig-and-regress point estimation method proceeds
in two steps:
\begin{lyxlist}{00.00.0000}
\item [{\textbf{Krig}}] Produce Kriging estimates of missing measurements
\[
\ensuremath{\hat{\mathbf{R}}=\hat{\mathbf{m}}+\bar{\mathbf{K}}_{\hat{\theta}}^{T}\mathbf{K}_{\hat{\theta}}^{*-1}\left(\mathbf{R}^{*}-\hat{\mathbf{m}}^{*}\right)},
\]
where $\hat{\theta}$ is an estimate of $\theta$, typically a maximum
likelihood estimate.
\item [{\textbf{Regress}}] Estimate (1) where $\mathbf{R}_{\mathrm{true}}$
is replaced with $\hat{\mathbf{R}}$, i.e., 
\[
\left(\begin{array}{c}
\hat{\beta}\\
\hat{\gamma}
\end{array}\right)=\left(\left(\hat{\mathbf{R}},F\right)^{T}\left(\hat{\mathbf{R}},F\right)\right)^{-1}\left(\hat{\mathbf{R}},F\right)^{T}\mathbf{Y}.
\]
\end{lyxlist}

Correct inference requires standard errors that take
into account the uncertainty brought about by the imputation of the
missing covariates. That is, one needs a two-step estimator for which
neither estimation nor inference requires knowledge of $\Sigma$ and
whose standard errors account for the uncertainty in the first step.

Versions of this problem have come up in the literature under many
guises \citep[see, for instance,][]{pagan1984econometric}. A very general case is addressed
by \citet{murphy1985estimation} who provide an asymptotic covariance formula
with a positive-definite correction term accounting for the variation
due to the estimation of the imputed regressors. However, such two-step
standard errors again require the stipulation of the covariance of
the regression errors, $\Sigma$.\footnote{There is a delicate conceptual point here. In the frequentist framework
with a DGP corresponding to \eqref{eq:varianceobservable}, the variance of the Krig-and-regress
estimate $\hat{\beta}_{KR}$, obtained by generating regressors $\hat{R}^{*}$
via Kriging and then regressing $Y$ on $\hat{R}^{*}$, depends on
$\Sigma$. Consequently, we must entertain a different but still credible
data generating process (DGP) in order to produce pivotal inference
methodology.}

\citet{maccini2009under} circumvent this issue by relying on a two-stage
least-squares approach. However, identification under this approach
is hard to argue, and estimates of the regression coefficients need
not be consistent (see Subsection \ref{ssec:maccinireg}). 

\citet{madsen2008regression} work out standard errors for the regression coefficient
estimated using the Krig-and-regress method. They provide a protocol
for estimating the unconditional (on the realization of the random
field for the misaligned regressor) variance of the regression coefficient.
They find, in their application, that the produced unconditional standard
errors differ only mildly from the OLS standard errors. Crucially,
the standard errors they provide do not account for the estimation
of the covariance parameter for rainfall, i.e., the uncertainty due
to first-stage estimation. 

Our concern, in contrast to that of \citet{madsen2008regression}, is to provide
confidence intervals that take into account the uncertainty due to
the estimation of the imputed regressor. Note that in the Krig-and-regress
method, accounting for the variation due to the estimation of the
imputed regressor is tantamount to accounting for the variance due
to the estimation of the covariance and mean parameters of the random
field of the misaligned regressor (say, rainfall).

Since the motivation for using a two-step method is to avoid the modeling
burden of specifying the covariance structure for the regression errors,
we  ought to produce standard errors that do not require evaluating
$\Sigma$. 

It is plausible that the residual errors of the best linear predictor
\[
Y(X)-E^{*}[Y(X)|R(X)],
\]
where $E^{*}$ is the best linear prediction operator, are spatially
correlated (for instance, through an omitted variable such as pollution,
which concentrates differently in different areas). Uncertainty assessments
of the Krig-and-regress coefficient estimates, if we do not condition
on the realization of $R$ and $Y$, are thus bound to rely on the
estimation of $\Sigma$.\footnote{Unless the estimation is done with standard errors so conservative that they are useless
in practice.} A different yet plausible stochastic model must thus be entertained.

\subsubsection{Survey Sampling}

Consider the identification strategy in the context of our main application.
Let $\mathscr{D}\subset\mathbb{R}^{2}$ be the geographic domain under
study. Let $R$ and $Y$ be the random fields for rainfall and the
outcome variable from a geolocated household survey (say, height),
respectively. Let $X^{*}\subset\mathscr{D}$ be the locations of the
rainfall measurement stations. Let $\hat{R}=E^{*}[R|R^{*}]$, where
$E^{*}$ is the best linear prediction operator and $R^{*}=R(X^{*})$
is the observed rainfall. Let $X\subset\mathscr{D}$ be the locations
of surveyed households. 

The key observation is that if all the uncertainty in the second step
arises from the resampling (with replacement) of the surveyed households,
then the observations $(Y(X_{i}),\hat{R}(X_{i}))$ are independently
distributed, conditional on the realization of $R$ and $Y$. That
is, conditional on the realization of the random field of rainfall
and the outcome variable at all households (but unconditional on which
household is randomly drawn with replacement from the population),
the observations are independent and identically distributed. 

Remarkably, by modeling the sources of randomness as detailed in the
previous paragraph, one can provide correct inference without relying
on knowledge of $\Sigma$. This is, of course, in contrast to standard
frequentist inference, which is unconditional on the observed data
and where uncertainty in the regression coefficients is thought to
capture the variation arising from repeated samples of $(Y,R,X)$.\footnote{For simplicity of exposition, we omit the additional covariates in
this section.} 

Indeed, we can treat the problem as one of survey sampling and estimate
the linear regression coefficient of $Y(\chi_{\mathrm{hh}})$ on $R(\chi_{\mathrm{hh}})$
over the whole population (say, of Indonesian households) for the
given realizations of the random fields $Y$ and $R$, where $\mathcal{X}_{\mathrm{hh}}$
is the set of locations of all households.

If the surveyed households are drawn with replacement from the full
population, then the corresponding observations $(Y_{i},R_{i})$ sampled
with replacement from $(Y(\chi_{\mathrm{hh}}),R(\chi_{\mathrm{hh}}))$
will be independent and identically distributed.

Crucially, this modeling approach can furthermore accommodate inference which takes into account
first-stage uncertainty. Indeed, variance due to estimation in the
first stage is captured by allowing the locations of the rainfall
stations to be random, while still conditioning on the realization
of the random field of rainfall.

We need to be clear as to what kind of process we have in mind to
accommodate the misaligned case. The variation in the outcome variable
comes from the random selection of the survey households, whose locations
are collected in $\mathbf{X}_{\mathrm{hh}}\subset\mathcal{X}_{\mathrm{hh}}$.
This is a natural assumption as it mimics the original data collection
process of the main application \citep{maccini2009under}, which was
itself a survey. We assume a similar survey sampling scheme of the
locations $\mathbf{X}_{\mathrm{rain}}$ of the rainfall stations from
the set of all possible rainfall locations, $\chi_{\mathrm{rain}}$.
We consider that $\chi_{\mathrm{rain}}$ contains all the locations
where the weather stations could have been and that their locations
collected in $\mathbf{X}_{\mathrm{rain}}$ were selected independently
and uniformly at random from $\chi_{\mathrm{rain}}$; i.e., we condition
on $R(\chi_{\mathrm{rain}})$ and it is the location of the weather
stations which is different if we resample a new data set from the
DGP.\footnote{Sampling of the weather station locations need not be uniform or over
a finite set; pivotal inference obtains as long as we condition on
$R(\cdot)$, and locations may be sampled according to any distribution
$f$ over the domain.}

The key point is that, since the sampling of the locations of the
realizations of $Y$ and $R$ are independent, the outcome data variation
does not inform the variation of the interpolated values. 

The exact nature of the target parameters depends on our conceptualization
of the DGP. We may still believe that the data arose from a law under
which the best linear predictor in population has the form $E^{*}[Y|R]=\beta_{0}+R\beta$,
and carry out inference conditional on the realization of the random
fields $Y(\cdot)$ and $R(\cdot)$. Of course, the target, conditional
on the realizations of the random fields, is the regression coefficient
of $Y(\chi_{\mathrm{hh}})$ on $R(\chi_{\mathrm{hh}})$, which we
call $\beta_{\chi_{\mathrm{hh}}}$.

However, under large domain asymptotics and standard ergodicity assumptions,
we consistently estimate the population best linear predictor coefficient,
$\beta$. The intuition for why conditional estimation delivers a
consistent estimate of an unconditional quantity is straightforward.
We can think of chopping off from an arbitrarily large random field
an increasing number of themselves increasingly large and increasingly
pairwise distant random fields; these increasingly many random fields
will be asymptotically independent from each other, thereby delivering
independent replications of random fields drawn from the underlying
DGP. Therefore, estimated coefficients will benefit from the consistency
properties that obtain under usual frequentist asymptotics.

We believe the assumption of resampling with replacement is innocuous.
Certainly, sampling without replacement describes more accurately
the sampling protocol of the survey. Nevertheless, the survey size
is so small compared to the population size that both sampling methods
(with and without replacement) yield the same observables with high
probability.

\subsubsection{Two-step Bootstrap Implementation}\label{sssec:bootstrap}

We detail the estimator which consistently estimates the population
regression coefficient and does not require specification of $\Sigma$
for neither point estimation or inference. 

Given $\theta$, and conditional on the realization of the random
field of rainfall as well as the outcome variable for each household
of the population, $\hat{\beta}$ only depends on which households
are drawn (randomly, with replacement) to be part of the survey. This
variation is captured by resampling using the bootstrap. This naturally
suggests a two-step bootstrap procedure in which $\theta$ is first
drawn, accounting for variation in $R(\mathbf{X}_{\mathrm{rain}})$,
to determine $\hat{R}(\theta)$, thus capturing the uncertainty due
to the estimation of $\theta$.

Instead of bootstrapping the rainfall data and estimating multiple
times a maximum likelihood estimate for the coefficients of the covariance
function of rainfall, we rely on the heuristic large sample argument
described above (applied to the first step), and use the much more
convenient asymptotic distribution of the maximum likelihood estimator
$\hat{\theta}_{\mathrm{mle}}$, obtained from training only on rainfall
data.
The full procedure is described in pseudocode as follows:

For each $j=1,...,J,$

\begin{itemize}
\item Draw $\hat{\theta}^{(j)}\sim N(\hat{\theta}_{\mathrm{mle}},\widehat{V(\hat{\theta}_{\mathrm{mle}})})$,
i.e., from its asymptotic distribution (using only $\mathbf{R}^{*}$
as data)
\item Compute $\hat{\mathbf{R}}^{(j)}=\hat{\mathbf{R}}(\hat{\theta}^{(j)})=\hat{\mathbf{m}}+\bar{\mathbf{K}}_{\hat{\theta}^{(j)}}^{T}\mathbf{K}_{\hat{\theta}^{(j)}}^{*-1}\left(\mathbf{R}^{*}-\hat{\mathbf{m}}^{*}\right)$
\item Draw new data set $\mathscr{D}^{(j)}$ with replacement from $\left(\mathbf{Y},\hat{\mathbf{R}}^{(j)}\right)$
\item Calculate $\hat{\beta}^{(j)}$, the regression coefficient for the
data set $\mathscr{D}^{(j)}$
\end{itemize}
Quantiles from the set of bootstrap draws $\{\beta^{(j)}\}_{j=1,...,J}$
can be used to form confidence intervals, and the average can be used
to give a point estimate.  See Section \ref{app:bootstrap} of the Appendix for the case of random fields $R(\cdot)$ with mean component $s(\cdot)^{T}\rho$.

\subsection{Minimum-Distance Estimation}

A simple identification argument naturally suggests a procedure delivering
pivotal point estimation. Upon inspection of \eqref{eq:variance}, we find that $\beta$
is identified from $V_{R^{*}}=\mathbf{K}^{*}$ and $V_{YR^{*}}=\beta\bar{\mathbf{K}}^{T}$
alone, and hence identification does not require modeling the covariance
structure of the regression errors. Specifically, using rainfall data
only, we obtain an estimate of the covariance parameter $\theta$
and thus obtain an estimate of $\bar{\mathbf{K}}$. Since $\beta\bar{\mathbf{K}}$
is directly identified from the covariance between the outcome data
and the observed rain, $\beta$ is identified. This naturally invites
a procedure that will rely on this identification observation to produce
a robust estimate of $\beta$.

We develop a minimum-distance estimator. As opposed to two-step methods,
all parameters showing up both in the first and second stage problems
are estimated once, jointly on both first and second stage data. In
order to conduct inference, we develop limit distribution theory,
which is detailed in Appendix \ref{app:distribution}. The resulting asymptotic covariance
matrix depends on the covariance of the regression errors, which may
be unpalatable to some users. In order to carry out pivotal inference,
we make a novel use of likelihood-free inference methods.

We state the minimum-distance estimator and its limit distribution
in terms of variograms to dovetail the results of \citet{lahiri2002asymptotic}.
Note that the minimum-distance estimator and theory articulated in
terms of covariances instead of variances obtains analogously (see,
for instance, Subsection \ref{sssec:pivotalinference}).

Let $\gamma_{R^{*}}(h;\phi)=V_{\phi}(R(x)-R(x+h))$ and $\gamma_{YR^{*}}(h;\phi)=V_{\phi}(R(x)-Y(x+h))$,
where $\phi=(\beta,\theta)$, be the variogram of $R^{*}$ and the
covariogram of $Y$ with $R^{*}$, respectively. Note that $\gamma_{YR^{*}}(h;\phi)=(1+\beta^{2})V_{\phi}\left(R(x)\right)-2\beta Cov_{\phi}\left(R(x),R(x+h)\right)$.
Let $\hat{\gamma}_{R^{*}}(h)$ and $\hat{\gamma}_{YR^{*}}(h)$ be
 nonparametric estimators of $\gamma_{R^{*}}(h;\phi)$ and
$\gamma_{YR^{*}}(h;\phi)$, respectively. Note that the method defined
in terms of covariances can be implemented analogously.

Let $\{h_{1},...,h_{K_{R^{*}}}\}$ and $\{c_{1},...,c_{K_{YR^{*}}}\}$
be finite sets of lag vectors in $\mathbb{R}^{2}$ such that $\hat{\gamma}_{R^{*}}(h_{i})$
is defined for all $i=1,...,K_{R^{*}}$ and $\hat{\gamma}_{YR^{*}}(c_{j})$
is defined for all $j=1,...,K_{YR^{*}}$.\footnote{The lags can be the default lags of the directional empirical variogram
estimator from a geostatistical package, such as $\mathrm{\mathtt{gstat}}$
in R.} Let 
\[
g_{n}(\phi)=\left(\hat{\gamma}_{YR^{*}}(c_{1})-\gamma_{YR^{*}}(c_{1};\phi),\dots,\hat{\gamma}_{YR^{*}}(c_{K_{YR^{*}}})-\gamma_{YR^{*}}(c_{K_{YR^{*}}};\phi),\right.
\]
\[
\left.\hat{\gamma}_{R^{*}}(h_{1})-\gamma_{R^{*}}(h_{1};\phi),\dots,\hat{\gamma}_{R^{*}}(h_{K_{R^{*}}})-\gamma_{R^{*}}(h_{K_{R^{*}}};\phi)\right)^{T}.
\]
For some positive-definite weighting matrix $B_{n}$, define the minimum-distance
estimator

\[
\hat{\phi}_{\text{M-D}}=\arg\min_{\phi\in\Phi}g_{n}(\phi)^{T}B_{n}g_{n}(\phi),
\]
for a convex support $\Phi=\mathbb{R}\times\Theta$. Then $\hat{\beta}_{\text{M-D}}$,
the estimate of $\beta$, does not depend on any specification of
the covariance structure of the regression errors. Different choices
of $B_{n}$ will correspond to different traditional estimators; $B_{n}=I$
yields the ordinary least-squares estimator, $B_{n}=\mathrm{diag}(b_{n,1}(\phi),...,b_{n,K_{YR^{*}}+K_{R^{*}}}(\phi))$,
for some choice of weights $b_{n,i}$, $i=1,...,K$, gives the weighted
least squares estimator, and $B_{n}(\phi)=\Sigma_{g}^{-1}(\phi)$,
where $\Sigma_{g}(\phi)$ is the asymptotic covariance matrix of $g_{n}(\phi)$
is the generalized least-square version of the minimum-distance estimator.
Our suggested rule of thumb is to use the efficient $B_{n}(\phi)=\Sigma_{g}^{-1}(\phi)$
evaluated at the Krig-and-regress estimates.

Another attractive feature of this estimator is its flexibility. The
vector of moments can be extended to accommodate other conditions,
perhaps motivated by economic theory. 

\subsubsection{Limit Distribution Theory for the Minimum-Distance Estimator}

In order to carry out inference using the proposed minimum-distance
estimator, we need asymptotic distribution theory for the statistic,
which is the empirical variogram (defined below). \citet{lahiri2002asymptotic}
develop such theory for data on a regular lattice, and give the asymptotic
distribution of the minimum-distance estimator as a corollary. \citet{lahiri2003central} proves a series of useful central limit theorems for spatial
statistics, some of which can be leveraged to extend the asymptotic
theory for the empirical variogram to the case of irregular data.
In contrast to the inferential framework of Section \ref{ssec:krigregress}, the inference here does not rely on survey sampling and is unconditional on the realization of the random fields.  Intuitively, the asymptotic framework considers the realization of a single ``arbitrarily large'' random field, and sampling variation obtains because arbitrarily distant observations are effectively independent.

To approximate the variogram
$E\left[\left(\varepsilon(x)-\varepsilon(x+h)\right)^{2}\right]$ of a given random
field $\varepsilon$, we define the empirical variogram \citep[p. 34]{gelfand2010handbook}, 
\[
\hat{\gamma}(h)=\frac{1}{\left|N_{n}(h)\right|}\sum_{(s_{i},s_{j})\in N_{n}(h)}\left(\hat{\varepsilon}(s_{i})-\hat{\varepsilon}(s_{j})\right)^{2},
\]
where $\hat{\varepsilon}(s)$ is an estimate of the random component
$\varepsilon(s)$ and the bin $N_{n}(h)$ is the set of pairs of observations
separated by a vector close to $h\in\mathbb{R}^{2}$. If, instead,
the summation is over $\left(\hat{\varepsilon}(s_{i})-\hat{\varepsilon}'(s_{j})\right)^{2}$
for distinct random fields $\varepsilon$ and $\varepsilon'$, then
we speak of the covariogram between both random fields and of its
empirical estimator.

The limit distribution theory for $g_{n}$ and $\hat{\phi}_{\text{M-D}}$
can be obtained in the pure- and mixed-increasing domains with the
so-called stochastic design \citep{lahiri2003central}. Explicitly, the sampling
region, denoted $\mathcal{R}_{n}$, is for each $n$ a multiple of
a prototype region $\mathcal{R}_{0}$, defined as follows. The prototype
region satisfies $\mathcal{R}_{0}^{*}\subset\mathcal{R}_{0}\subset\bar{\mathcal{R}}_{0}^{*}$,
where $\mathcal{R}_{0}^{*}$ is an open connected subset of $(-1/2,1/2]^{2}$
containing the origin. Let $\left\{ \lambda_{n}\right\} _{n\in\mathbb{N}}$
be a sequence of positive real numbers such that $n^{\epsilon}/\lambda_{n}\rightarrow0$
as $n\rightarrow\infty$ for some $\epsilon>0$. Then the sampling
region is defined as
\[
\mathcal{R}_{n}=\lambda_{n}\mathcal{R}_{0}.
\]
To avoid pathological cases, we will assume that the boundary of $\mathcal{R}_{0}$
is delineated by a smooth function. This assumption can be modified
and weakened to adapt to other domains (e.g., star-shaped), see \citet{lahiri2003central}.

Furthermore, we speak of a stochastic design because the data is not
placed on a regular lattice, and observation locations must be modeled
otherwise. They are modeled as follows. Let $f(x)$ be a continuous,
everywhere positive density on $\mathcal{R}_{0}$, and let $\{X_{n}\}_{n}$
be a sequence of independent and identically distributed draws from
$f$. Let $x_{1},...,x_{n}$ be realizations of $X_{1},...,X_{n}$,
and define the locations $s_{1},...,s_{n}$ of the observed data in
$\mathcal{R}_{n}$ as
\[
s_{i}=\lambda_{n}x_{i},\ i=1,...,n.
\]

In the stochastic design, pure-increasing asymptotics require that
$n/\lambda_{n}^{2}\rightarrow C$ for some $C\in(0,\infty)$ as $n\rightarrow\infty$. Mixed-increasing asymptotics require that $n/\lambda_{n}^{2}\rightarrow\infty$.

First, we obtain a central limit theorem for the statistic entering
the minimum-distance objective function. For simplicity of exposition,
take $g_{n}$ as defined above but in which enter as statistics only
variograms in $R$, and let $\{h_{1},...,h_{K_{R^{*}}}\}$, $K_{R^{*}}\in\mathbb{N}$,
be the full set of lag vectors. The result is trivially extended to
accommodate as statistics the covariogram in $R$ and $Y$. 

Define the mixing coefficient
\[
\alpha(a;b)=\sup\left\{ \tilde{\alpha}(T_{1},T_{2}):d(T_{1},T_{2})\geq a,\ T_{1},T_{2}\in\mathcal{S}_{3}(b)\right\} ,
\]
where $\mathcal{S}_{3}(b)=\left\{ \bigcup_{i=1}^{3}D_{i}:\sum_{i=1}^{3}\left|D_{i}\right|\le b\right\} $
is a collection of disjoint unions of three cubes $D_{1},D_{2},D_{3}$
in $\mathbb{R}^{2}$, $d(T_{1},T_{2})=\min\left\{ \left\Vert x_{1}-x_{2}\right\Vert :x_{1}\in T_{1},x_{2}\in T_{2}\right\} $,
\[
\tilde{\alpha}(T_{1},T_{2})=\sup\left\{ \left|P\left(A\cap B\right)-P(A)P(B)\right|:A\in\sigma\left\langle R(s)|s\in T_{1}\right\rangle ,B\in\sigma\left\langle R(s)|s\in T_{2}\right\rangle \right\} ,
\]
and  $\sigma\left\langle R(s)|s\in T\right\rangle $ is the $\sigma$-field
generated by the variables $\left\{ R(s)|s\in T\right\} $, $T\subset\mathbb{R}^{2}$.

Suppose there exists a non-increasing function $\alpha_{1}(\cdot)$
with $\lim_{a\rightarrow\infty}\alpha_{1}(a)=0$ and a non-decreasing
function $g(\cdot)$ such that
\[
\alpha(a,b)\le\alpha_{1}(a)g(b),\ a>0,b>0.
\]

Consider $N_{u,n}(h_{k})=\left\{ (i,j)\in N_{n}(h_{k}):j\le j'\ \mathrm{for\ all}\ (i,j')\in N_{n}(h_{k})\right\} $,
a set of ordered pairs in $N_{n}(h_{k})$ with unique starting locations
$i$, and define the difference sets $N_{r,n}(h_{k})=N_{n}(h_{k})\backslash N_{u,n}(h_{k})$
and $N'(h_{k})=\{1,...,n\}\backslash\left\{ i:(i,j)\in N_{u,n}(h_{k})\ \mathrm{for\ some\ }j\right\} $.  We give theory for two-dimensional random field with unknown constant mean.  The general case is stated and proved in the Appendix.

\begin{theorem}
Suppose that $\left\{ \varepsilon(x):x\in\mathbb{R}^{2}\right\} $
is a stationary random field
such that $E\left|\varepsilon(0)\right|^{2+\delta}<\infty$
for some $\delta>0$. Suppose $f$ is continuous and everywhere positive
on $\overline{\mathcal{R}}_{0}$, and that $\int_{\mathcal{R}_{0}}f^{2}(x)dx<\infty$.
Let $\alpha_{1}(a)=a^{-\tau}$ for some $\tau>\frac{2(2+\delta)}{\delta}$
and suppose $g(b)=o\left(b^{\frac{\tau-2}{8}}\right)$. Suppose that
$\left(\log n\right)^{2}\lambda_{n}^{\frac{2-\tau}{4\tau}}\rightarrow0$
as $n\rightarrow\infty$. Further suppose that the autocovariance function
$\sigma_{ij}(x)=Cov_{\phi_{0}}\left(\left(\varepsilon(0)-\varepsilon(h_{i})\right)^{2},\left(\varepsilon(x)-\varepsilon(x+h_{j})\right)^{2}\right)$
satisfies $\int\left|\sigma_{ij}(x)\right|dx<\infty$, $i,j=1,...,K$.
Suppose that $\lambda_{n}^{2}\left\Vert \hat{m}-m\right\Vert _{2}^{4}=o_{p}(1)$,
$\left|N_{n}(h_{k})\right|=\left(1+o(1)\right)n$, $\left|N_{r,n}(h_{k})\right|=o(n^{\frac{1}{2}})$,
$\left|N'_{n}(h_{k})\right|=o(n^{\frac{1}{2}})$, and $E\left[\left|\left(\varepsilon(x_{i})-\varepsilon(x_{j})\right)^{2}-\left(\varepsilon(x_{i})-\varepsilon(x_{i}+h_{k})\right)^{2}\right|\right]=o(n^{-\frac{1}{2}})$
for all $(i,j)\in N_{n}(h_{k})$, $k=1,...,K$.
\renewcommand{\labelenumi}{(\roman{enumi})}
\begin{enumerate}
\item If $n/\lambda_{n}^{2}\rightarrow C_{1}\in(0,\infty)$ as
$n\rightarrow\infty$, then
\[
n^{\frac{1}{2}}g_{n}(\phi_{0})\overset{d}{\rightarrow}N\left(0,\Sigma_{g}(\phi_{0})\right), \quad a.s.\ P_{X}, 
\]
where the $i,j$ entry of the covariance matrix is
$\left(\Sigma_{g}(\phi_{0})\right)_{ij}=\sigma_{ij}(0)+Q\cdot C_{1}\cdot\int_{\mathbb{R}^{2}}\sigma_{ij}(x)dx$,
with $Q=\int_{\mathcal{R}_{0}}f^{2}(x)dx$.
\item If $n/\lambda_{n}^{2}\rightarrow\infty$ as $n\rightarrow\infty$,
then
\[
\lambda_{n}g_{n}(\phi_{0})\overset{d}{\rightarrow}N\left(0,\Sigma_{g}(\phi_{0})\right) \quad  a.s.\ P_{X},
\]
where $\left(\Sigma_{g}(\theta_{0})\right)_{ij}=Q\cdot\int_{\mathbb{R}^{2}}\sigma_{ij}(x)dx$.
\end{enumerate}
\label{thm:1}
\end{theorem}

The assumptions on the mixing rates are standard (see, for instance,
\citet{lahiri2002asymptotic} and \citet{lahiri2003central}). The assumption on the bin
accuracy $E\left[\left|(\varepsilon(s_{i})-\varepsilon(s_{j}))^{2}-(\varepsilon(s_{i})-\varepsilon(s_{i}+h_{k}))^{2}\right|\right]=o(n^{-1/2})$,
while it is realistic for applications such as ours where sampled locations are meant to be spread out somewhat evenly, is strong. It is, however, intrinsic
to the matter at hand; the bias in the moment condition must vanish
in the $\sqrt{n}$-asymptotics. This theoretical condition connects
with our practical experience in simulations; a careful choice of
lags for which even small bins $N_{n}(h)$ will contain a large enough
number of pairs makes for a noticeably more accurate asymptotic approximation
of the distribution. Further note that, as discussed in Section 3,
the coverage is close to nominal in simulation using real data.

With the limit distribution of the statistic in hand, the central
limit distribution of the minimum-distance estimator obtains under
additional identifying assumptions.
\begin{assumption}
Suppose that
\renewcommand{\labelenumi}{(\roman{enumi})}
\begin{enumerate}
    \item For any $\epsilon>0$, there exists $\nu>0$
such that $\inf\left\{ \sum_{i=1}^{K}\left(\gamma(h_{i};\theta)-\gamma(h_{i};\theta')\right)^{2}:\left\Vert \theta-\theta'\right\Vert \ge\epsilon\right\} >\nu$,
    \item $\sup\left\{ \gamma(h;\theta):h\in\mathbb{R}^{2},\theta\in\Theta\right\} <\infty$,
and $\gamma(h;\theta)$ is continuously differentiable in $\theta$,
    \item $B_{n}(\theta)$ is positive definite for
all $\theta\in\Theta$ and $\sup\left\{ \left\Vert B_{n}(\theta)\right\Vert +\left\Vert B_{n}(\theta)^{-1}\right\Vert :\theta\in\Theta\right\} <\infty$,
and $B_{n}(\theta)$ is continuously differentiable in $\theta$ for
all $n$.
\end{enumerate}
\label{as:1}
\end{assumption}
The main distributional result may now be stated. Let $g_{j}(\theta)$
be the gradient of $g_{n}(\theta)$ with respect to $j^{\mathrm{th}}$
coordinate of $\theta$.
\begin{corollary}
Suppose that the conditions stated in Theorem
\ref{thm:1} hold and that $\Sigma_{g}(\theta_{0})$ is positive
definite. Let $b_{n}=n/\lambda_{n}^{2}$. Then under the conditions
cited in Assumption \ref{as:1}, if the matrix of partial derivatives $\Gamma(\theta_{0})=\left(g_{1}(\theta_{0});...;g_{\dim(\Theta)}(\theta_{0})\right)$
is full rank,
\[
b_{n}n^{1/2}(\hat{\theta}_{n}-\theta_{0})\overset{d}{\rightarrow}N\left(0,\Sigma(\theta_{0})\right),
\]
where $\Sigma(\theta_{0})=A(\theta_{0})\Gamma(\theta_{0})^{T}B(\theta_{0})\Sigma_{g}(\theta_{0})B(\theta_{0})\Gamma(\theta_{0})A(\theta_{0})$,
and $A(\theta_{0})=\left(\Gamma(\theta_{0})^{T}B(\theta_{0})\Gamma(\theta_{0})\right)^{-1}$.
\label{cor:1}
\end{corollary}
The density of the observation locations $f$ has an intuitive impact
on the asymptotic covariance. As one would expect, if the observations
are well spread geographically, this makes for a lower variance because
$Q=\int_{R_{n}}f^{2}(x)dx$ is smaller. Correspondingly, cluttered
data arranged as a few clusters provides worse information, and the
variance is greater for it. Estimation of the asymptotic variance
is discussed in the Appendix.

\subsubsection{Pivotal Inference}\label{sssec:pivotalinference}

The asymptotic variance of the minimum-distance estimator depends
on the covariance matrix of regression errors, $\Sigma$. That is,
point estimation is pivotal with respect to $\Sigma$, but large sample
inference is not. The reason for this is fairly intuitive. The stability
of the point estimate is informed by variability of $R$, which is
itself informed by the variability of $Y$, which can only be recovered
if it is disentangled from the variability of the error term.

The pivotal point estimation motivates the use of a Monte Carlo sampler
using the minimum-distance objective as a basis for its acceptance
criteria. Specifically, we want a Monte Carlo procedure that samples
a proposed coefficient $\phi^{*}=(\beta^{*},\theta^{*})$ when moments
estimated or simulated at $\phi^{*}$ are close to their observed,
empirical counterpart. Two approaches naturally suggest themselves:
an approximate Bayesian computation (ABC) approach \citep{forneron2018abc} and a Metropolis-Hastings approach \citep{chernozhukov2003mcmc}.

Typically, ABC is used as a likelihood-free approach to inference
when the likelihood, even up to a constant of proportionality (i.e.,
even without computing the normalizing constant), is computationally
intractable. In this case, however, ABC comes in handy not because
it circumvents likelihood computations but because it allows us to
provide confidence intervals for the minimum-distance estimator that
do not depend on the covariance of regression errors, $\Sigma$.

We develop the ABC sampler for moments built with the sample covariances,
but the same approach can be employed with variograms at the cost
of estimating a few nuisance parameters.

Consider the following derivation of a large sample approximation
to the ABC procedure. For purposes of exposition, first consider in
isolation the cross-covariance term at lag $h_{k}$ for a given $k$.
The ABC procedure is trying to sample coefficients $\theta$, $m$
and $\beta$ that make differences such as 
\[
\frac{1}{\left|N_{n}(h_{k})\right|}\sum_{(i,j)\in N_{n}(h_{k})}\left(R(s_{i})-\hat{m}\right)Y(s_{j})-\frac{1}{\left|N_{\tilde{n}}(h_{k})\right|}\sum_{(i,j)\in\tilde{N}_{\tilde{n}}(h_{k})}\left(\tilde{R}(s_{i})-\hat{\tilde{m}}\right)\tilde{Y}(s_{j})
\]
small, where $\tilde{R}$ and $\tilde{Y}$ are random fields generated
from a generative model evaluated at some candidate value $(\beta^{*},\theta^{*})$,
$\tilde{n}$ is the number of synthetic draws, and $\hat{m}$ and
$\hat{\tilde{m}}$ are the estimated means of rainfall in the true
and synthetic data sets, respectively.  The goal of the ABC procedure is thus to select coefficients such that synthetic data generated according to these coefficients makes the above difference between sample moments evaluated on true and synthetic data as small as possible.

\emph{A priori}, this approach may seem counterintuitive since generating synthetic
draws of $\tilde{Y}$ requires stipulating $\Sigma$. However, observing
that the moment estimate may be expressed as 
\[
\frac{1}{\left|N_{n}(h_{k})\right|}\sum_{(i,j)\in N_{n}(h_{k})}\left(R(s_{i})-\hat{m}\right)R(s_{j})\beta+\frac{1}{\left|N_{n}(h_{k})\right|}\sum_{(i,j)\in N_{n}(h_{k})}\left(R(s_{i})-\hat{m}\right)\varepsilon_{j}
\]
and that $\frac{1}{\left|N_{n}(h_{k})\right|}\sum_{(i,j)\in N_{n}(h_{k})}\left(R(s_{i})-\hat{m}\right)\varepsilon_{j}\rightarrow0$,
$\frac{1}{\left|N_{n}(h_{k})\right|}\sum_{(i,j)\in N_{n}(h_{k})}\left(R(s_{i})-\hat{m}\right)R(s_{j})\beta\rightarrow$\\
$\beta Cov_{\theta}(R(s),R(s+h_{k}))$
as $n\rightarrow\infty$, and $\frac{1}{\left|\tilde{N}_{\tilde{n}}(h_{k})\right|}\sum_{(i,j)\in\tilde{N}_{\tilde{n}}(h_{k})}\left(\tilde{R}(s_{i})-\hat{m}\right)\tilde{R}(s_{j})\beta^{*}\rightarrow\beta Cov_{\theta^{*}}(R(s),R(s+h_{k}))$
as $\tilde{n}\rightarrow\infty$, we may instead consider the difference
\[
\frac{1}{\left|N_{n}(h_{k})\right|}\sum_{(i,j)\in N_{n}(h_{k})}\left(R(s_{i})-\hat{m}\right)Y(s_{j})-\beta^{*}Cov_{\theta^{*}}(R(0),R(h_{k}))
\]
as a fit criterion. The full statistic is then a vector of such differences,
which we may again designate by $g_{n}$ as it is the same vector
statistic used for point estimation, when moments are built with covariances
as opposed to semivariances.

The full fit criteria is the quadratic form $l(\phi)=g_{n}(\phi)^{T}\tilde{B}_{n}g_{n}(\phi)$,
where $\phi=(\beta,\theta)$. As with point estimation, a positive-definite
weighting matrix $\tilde{B}_{n}$ must be used. A naive estimate of
the asymptotic variance -- using a covariance matrix proportional
to the identity for the regression errors -- performs well.
More in the spirit of inference by simulation using a generative model,
the inverse of an empirical estimate of the variance of $g_{n}(\phi)$,
evaluated on synthetic data from the generative model evaluated at
the Krig-and-regress estimates, has been found to work very well and
to be easy to implement -- and this estimate can likewise
be used in the evaluation of the covariance matrix of the proposal
distribution $q$ defined below.

The pseudocode for the procedure is as follows. Draw $\phi^{(0)}$
from $q$.\footnote{There is no burn-in period, but the sample starts at the first accepted
coefficient, $\phi^{(1)}$.} For $j=1,...,J$,
\begin{itemize}
\item Propose $\phi^{*}\sim q$
\item Draw a uniform random variable $u\sim U[0,1]$
\item Draw a sample
\begin{itemize}
\item if $u\le\mathbf{1}\left\{ l(\phi^{*})\le(1+\xi)\cdot l(\hat{\phi})\right\} \cdot \frac{q(\phi^{(j-1)})}{q(\phi^{*})} $,
then $\phi^{(j)}=\phi^{*}$
\item otherwise, $\phi^{(j)}=\phi^{(j-1)}$
\end{itemize}
\end{itemize}
where $q$ is a proposal distribution, which we pick to be a naive
approximation\footnote{We propose $\phi^{*}$ according to its two-step bootstrap distribution.} to the asymptotic distribution of $\hat{\phi}$, and
$\xi$ is a tolerance parameter. In the companion R package SpReg, the default setting is $\xi=0.1$, but it is best to tune the threshold $\xi$ to ameliorate mixing
and by inspection of the Monte Carlo distribution $\left\{ \phi^{(1)},...\phi^{(J)}\right\} $,
the acceptance ratio, and the trace plots.\footnote{We do not develop on MCMC diagnostics here, but refer the interested
reader to \citet{hoff2009first}, \citet{gelman2013bayesian}, and \citet{pouliot2020lecture}.}

As detailed in Table \ref{tab:cv} of Section \ref{sec:madsen} below, the heuristic appears to
be reliable, slightly over-rejecting in semi-synthetic simulations.

The suggested ``asymptotic ABC'' is conceptually close to a Metropolis-Hastings
algorithm using the minimum-distance objective in lieu of a target/posterior
distribution. 
Alternatively, one could exponentiate the minimum-distance objective,
as suggested in \citet{chernozhukov2003mcmc}. Perhaps surprisingly,
this approach seems to perform worse; we obtained substantially worse
mixing and coverage for a comparable tuning effort. It has, however,
the advantage of not including a tolerance parameter such as $\xi$
for acceptance.

\section{Revisiting \citet{madsen2008regression}}\label{sec:madsen}

We assess the performance of our suggested methods as well as that
of competing approaches on real and hybrid data. 

We apply and compare the methods under study using the cross-validation
exercise of \citet{madsen2008regression}. We use the same data set\footnote{We  would like to thank the authors for kindly providing their data
for replication.} as in their article. As explained therein and further detailed in
\citet{herlihy1998relationship}, the data is a subset of the sample obtained
for the Environmental Monitoring and Assessment Program of the Environmental
Protection Agency. All samples are from rivers and streams in the
American Mid-Atlantic region, and the analysis objective was to relate
stream characteristics with land use variables. There are 558 observations
over an area of 400,000 squared kilometers. The outcome variables
$Y(x)$ are the logarithm of chloride concentration at locations $x$,
and the independent variables $R(x^{*})$ are the logit transformations
of the percent of watershed in forest at locations $x^{*}$.

The reference value is obtained by doing generalized least-squares
on the full, aligned data set weighting with an estimate of the regression
error covariance: we obtain $\hat{\beta}_{\mathrm{full}}=-0.38$.
The simulation is implemented as follows; for each run, a randomly
chosen half of the independent variables, the $R_{i}$'s, are ``hidden'',
and the outcome variables, the $Y_{i}$'s, are ``hidden'' for the
other half of the data set, thus creating a misaligned data set. For
each round of the cross-validation exercise, $\beta$ is estimated
with each method, and estimates are recorded. 

\begin{figure}
\begin{center}\includegraphics[scale=0.35]{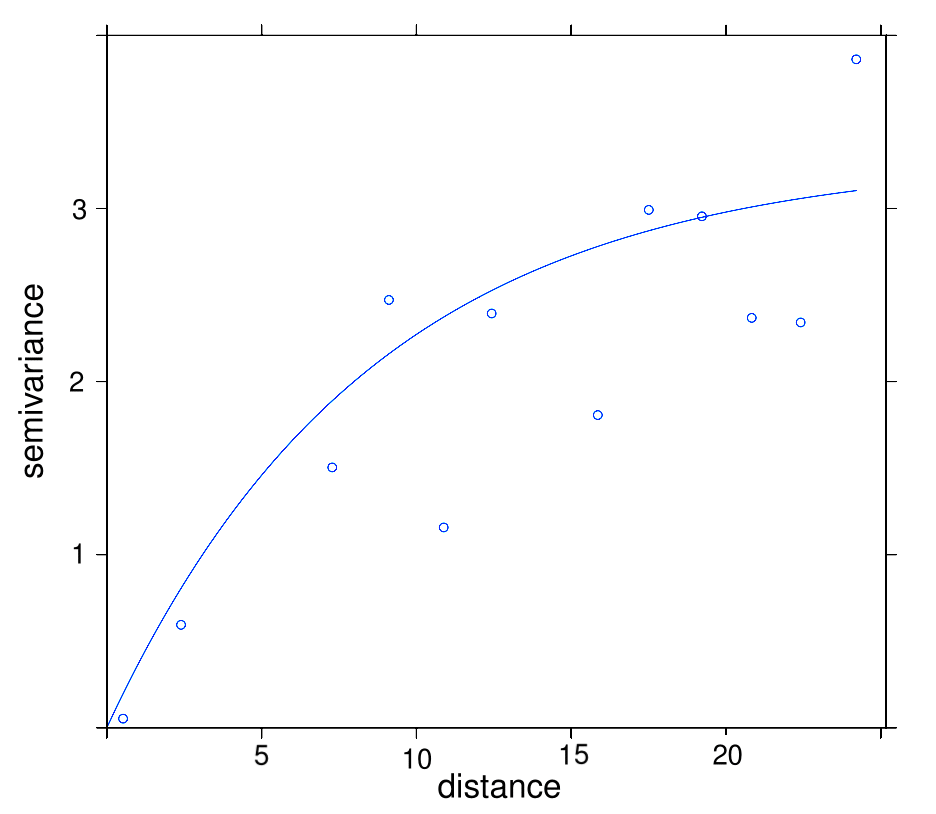}\end{center}

\caption{Empirical and fitted variogram of $R$. }
\label{fig:variogram}
\end{figure}



\begin{table}

\centering\footnotesize
\renewcommand{\arraystretch}{1.5}
\begin{tabular}{cccccccc}
\toprule 
{Estimation} & {Inference} & {$E[\hat{\beta}]$} & {RMSE($\hat{\beta}$)} & {$\sqrt{V(\hat{\beta})}$} & {$\sqrt{\hat{V}(\hat{\beta})}$} & {RMSE$\left(\sqrt{V(\hat{\beta})}\right)$} & {Coverage}\\
\midrule
{1-NN-and-Regress} & {naive} & {0.8642} & {0.1488} & {0.0608} & {0.0448} & {0.0162} & {0.23}\\
{4-NN-and-Regress} & {naive} & {1.0142} & {0.1099} & {0.1092} & {0.0735} & {0.0361} & {0.82}\\
\cmidrule{1-2}
\multirow{2}{*}{Krig-and-regress} & {naive} & {0.9947} & {0.0715} & {0.0717} & {0.0509} & {0.0209} & {0.84}\\
 & {2S Bootstrap} & {0.9947} & {0.0715} & {0.0717} & {0.0948} & {0.0505} & {0.95}\tabularnewline
\cmidrule{1-2} 
\multirow{2}{*}{Min Dist} & {Large Sample} & {1.0006} & {0.0636} & {0.0637} & {0.0565} & {0.0074} & {0.92}\\
 & {ABC} & {1.0310} & {0.2039} & {0.2026} & {0.1548} & {0.0864} & {0.91}\\
\bottomrule
\end{tabular}

\caption{Misaligned regression on lattice.}
\label{tab:lattice}
\end{table}

We see from the empirical variogram displayed in Figure \ref{fig:variogram} that neighboring
dependent variables do covary, thus allowing for useful interpolation
as a first step. 

\begin{figure}
\begin{center}\includegraphics[scale=0.2]{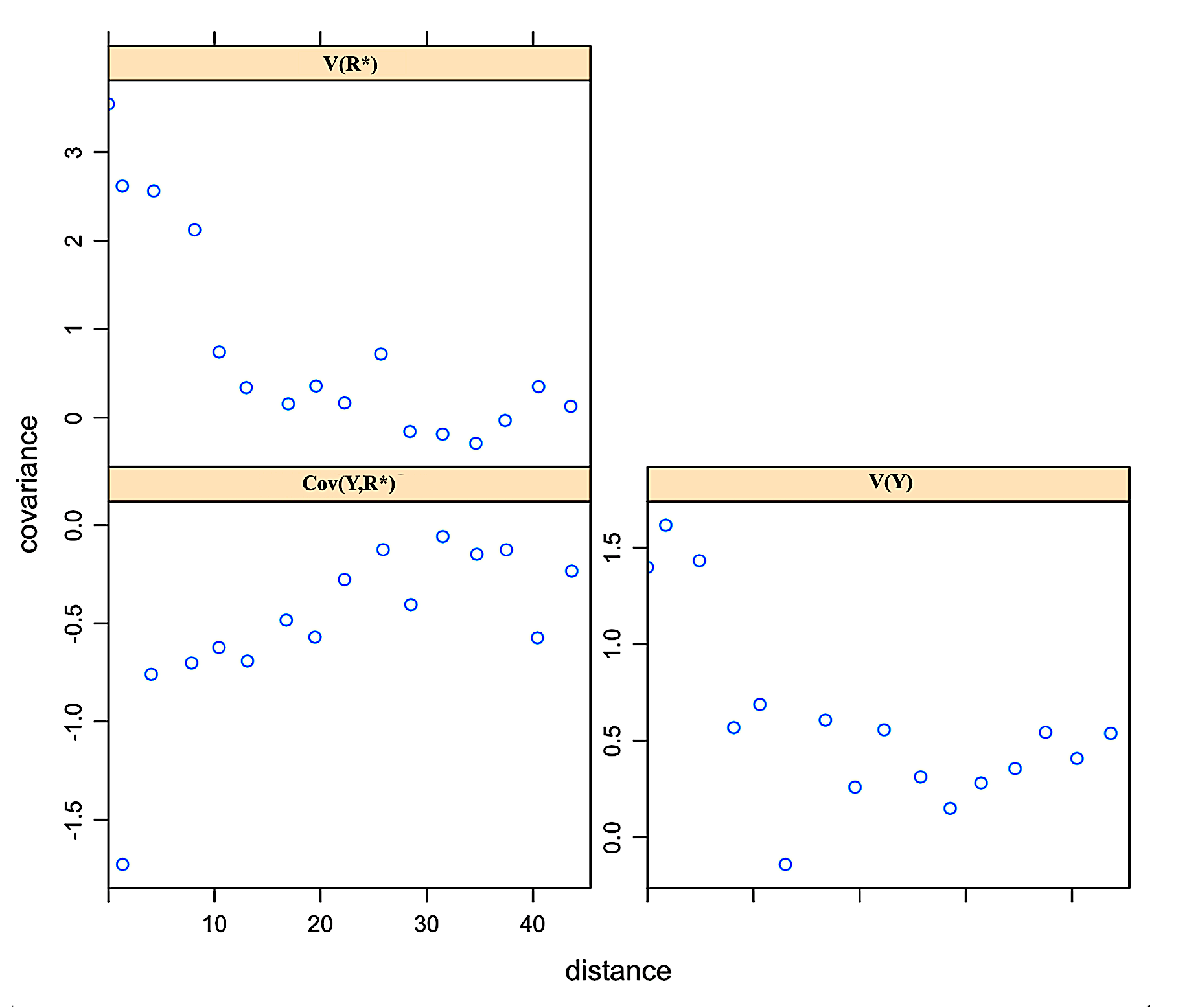}\end{center}

\caption{Nonparametric estimates of semivariance and covariance for $Y$ and $R^{*}$
as a function of distance. }
\label{fig:nonparam}
\end{figure}



Table \ref{tab:lattice} presents hybrid data on a regular lattice. This special case
is important as this is a common design. Rainfall is generated according
to the exponential covariance model with parameters estimated on the aligned
rivers data, but data is generated at locations on a lattice instead of the original, irregular locations. The outcome variable is generated according to the linear
regression model with parameters estimated on the aligned rivers data.
The mean $E[\hat{\beta}]$ is the average point estimate over all
simulation draws. The simulation standard errors $\sqrt{V(\hat{\beta})}$
are the standard deviations of the simulation draws, and the output
standard errors $\sqrt{\hat{V}(\hat{\beta})}$ give the average standard
errors over all simulation draws. The squared-root mean-squared errors
(RMSE) are defined analogously. For the two-step bootstrap, the mean
is the average mean over the bootstrap samples, and the regression
output standard errors are given by the average over simulation runs
of the standard deviation of bootstrap estimates. The coverage corresponds
to the fraction of times the confidence interval, computed with the
output standard errors of the current run of the simulation, covered
-0.38.

On the lattice, as detailed in Table \ref{tab:lattice}, the efficiency gains of the
minimum-distance approach deliver the expected improvement in squared-root
mean-squared error (RMSE) of the regression coefficient estimate.
The Krig-and-regress estimator, with naive inference ignoring first-stage
uncertainty, severely undercovers while the two-step bootstrap estimator
has correct coverage.

\begin{table}

\centering\footnotesize
\renewcommand{\arraystretch}{1.5}
\begin{tabular}{cccccccc}
\toprule 
{Estimation} & {Inference} & {$E[\hat{\beta}]$} & {RMSE($\hat{\beta}$)} & {$\sqrt{V(\hat{\beta})}$} & {$\sqrt{\hat{V}(\hat{\beta})}$} & {RMSE$\left(\sqrt{V(\hat{\beta})}\right)$} & {Coverage}\\
\midrule
{1-NN-and-Regress} & {naive} & {-0.168} & {0.207} & {0.033} & {0.033} & {0.002} & {0}\\
{4-NN-and-Regress} & {naive} & {-0.332} & {0.071} & {0.054} & {0.052} & {0.004} & {0.86}\\
\cmidrule{1-2}
\multirow{2}{*}{Krig-and-regress} & {naive} & {-0.395} & {0.063} & {0.060} & {0.063} & {0.007} & {0.94}\\
 & {2S Bootstrap} & {-0.395} & {0.063} & {0.060} & {0.064} & {0.012} & {0.95}\\
 \cmidrule{1-2}
\multirow{2}{*}{Min Dist} & {Large Sample} & {-0.421} & {0.109} & {0.098} & {0.114} & {0.028} & {0.96}\\
 & {ABC} & {-0.400} & {0.101} & {0.095} & {0.106} & {0.045} & {0.90}\tabularnewline
\bottomrule 
\end{tabular}

\caption{Cross-validation exercise using real data.}
\label{tab:cv}
\end{table}

Table \ref{tab:cv} presents the output of the cross-validation exercise using
the observed data. While both the Krig-and-regress
and the minimum-distance estimator do well in terms of RMSE and coverage
accuracy, the Krig-and-regress estimator outperforms the Minimum-Distance
estimator in both. This is explained in part to the difficulty of estimating
the weighting matrix in such a small sample, which is important for
to the quality of point estimation with the minimum-distance estimator.\footnote{One avenue we have found fruitful for weighing matrix estimation in
small samples is to simulate data from the posited model evaluated
at the Krig-and-regress estimates, and compute the variance of synthetic
sample moments using the observed location and simulated observations.
We leave further exploration of this approach for follow-up research.}

As noted in Table \ref{tab:method}, estimation of the asymptotic variance of the
minimum-distance estimator requires stipulation of $\Sigma$. The
pivotal ABC alternative appears to do relatively well, even though
it over-rejects by about 5\% in our simulations, where the interval
has 95\% nominal coverage.

\section{Reanalysis of \citet{maccini2009under}}\label{sec:maccini}

In \emph{Under the Weather: Health, Schooling, and Economic Consequences
of Early-Life Rainfall}, \citet{maccini2009under} estimate the effect
of a rainfall shock in infancy on adult socioeconomic outcomes such
as education or health. The paper merges a rainfall measurements data
set with survey data, both of which have geographically located observations.
The data sets are misaligned, as can be seen from Figure \ref{fig:map}, which
plots the respective locations.

\subsection{Data Description}

The data set for the regression is obtained by merging two misaligned
data sets. The first data set contains rainfall measurements from
measuring stations across Indonesia. The whole rainfall data spans
the years from 1864 to 2004.

Only the years 1953-1975 are used to identify yearly variation; the
other years are used to estimate long term averages. In almost every
year used in the analysis, more than 300 rainfall measurement stations
are active. The rainfall data comes from the Global Historical Climatology
Network (GHCN), which is a publicly available data set.\footnote{Available online at http://www.ncdc.noaa.gov/oa/climate/research/ghcn/ghcn.html.}

The second data set is the third wave of the Indonesian Family Life
Survey (IFLS3), which includes each surveyed individual's year and
location of birth. It consists of 4,615 women and 4,277 men born outside
large cities between 1953 and 1974.

The locations of the birthplaces and rainfall stations are given in
Figure \ref{fig:map}. We can see that the data sets are misaligned. We can also
see from Figure \ref{fig:map} that most birthplaces are situated fairly close
to one or more rainfall stations; the median distance to the closest
rainfall station is 11.43 km, and the third quartile distance is 90.28
km. The median distance to the fifth closest station is 30.24 km,
and the third quartile distance to the fifth closest station is 317.10
km. We show below that, at those distances, the rainfall measurements
are still correlated; thus, informative interpolation is possible. 

\begin{figure}
\begin{center}\includegraphics[scale=0.2]{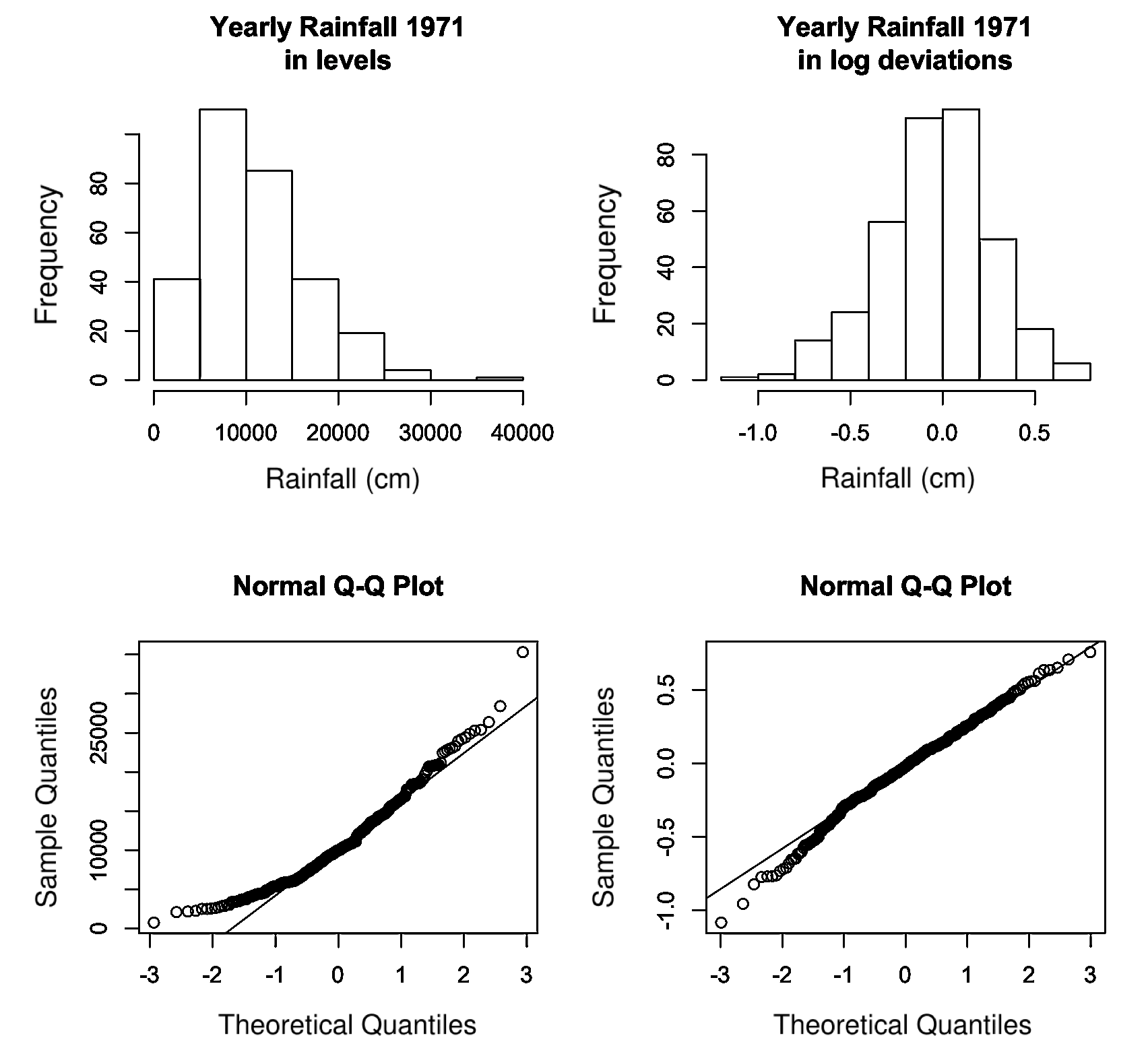}\end{center}

\caption{Histograms and Gaussian QQ-plots of 1971 rainfall in levels (left)
and in logarithm of ratio of yearly sum to long run yearly average
(right).}
\label{fig:histqq}
\end{figure}

We use, as \citet{maccini2009under} did, the log of the ratio of the
yearly rainfall with the long run average yearly rainfall. We find
in Figure \ref{fig:histqq} that this has the benefit of eliminating the point mass
at zero, and making the distribution ``closer to Gaussian.''

\begin{figure}
\begin{center}\includegraphics[scale=0.2]{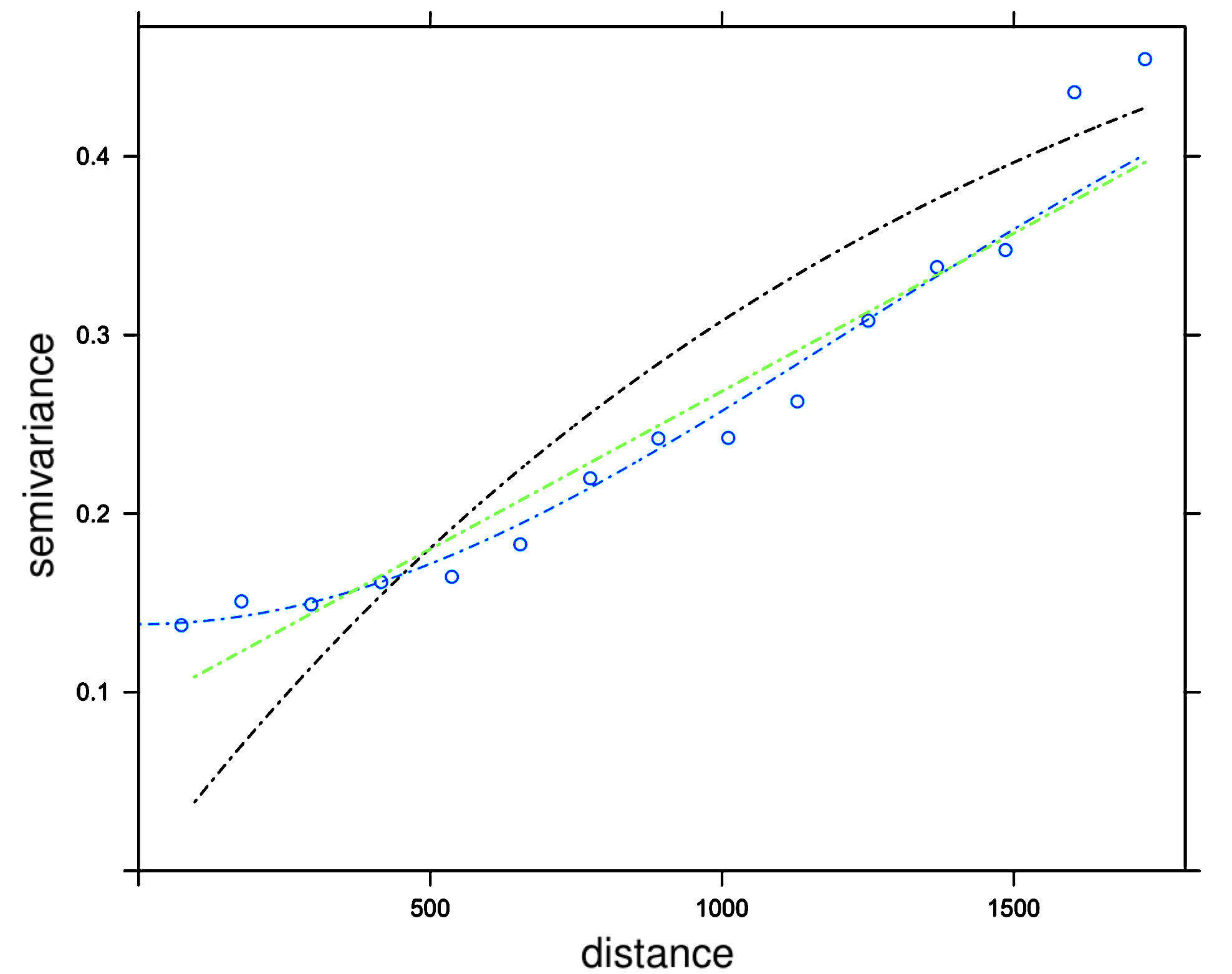}\end{center}

\caption{Plots of empirical and fitted semivariogram using the exponential
(black), Gaussian (blue) and linear (green) models. The data is from
the year 1971.}
\label{fig:semivariogram1971}
\end{figure}

The typical variogram fit of Figure \ref{fig:semivariogram1971} suggests a good fit of the variogram
and corroborates the assumption of a Gaussian covariance function
\eqref{eq:gaussiancov}. However, \citet{stein2012interpolation} warns against relying on such plots to
draw definitive conclusions, and likelihood fits were likewise investigated.
Altogether, this suggests that Kriging, which is the best
predictor under Gaussianity, ought to produce reliable
interpolation. 

\subsection{Regression Analysis}\label{ssec:maccinireg}

The first order of business is to implement the two-step estimator.
As mentioned above, we carry out Kriging on the log ratio of yearly
rainfall to long run yearly average. The transformed data has the
additional virtue of making immediate the comparison with the regression
output of \citet{maccini2009under}. Remark that if the imputed covariate
were a nonlinear transformation of the interpolated quantity, the estimated
coefficients may not be consistent estimators of the best linear predictor
coefficients.

\begin{figure}
\begin{center}\includegraphics[scale=0.25]{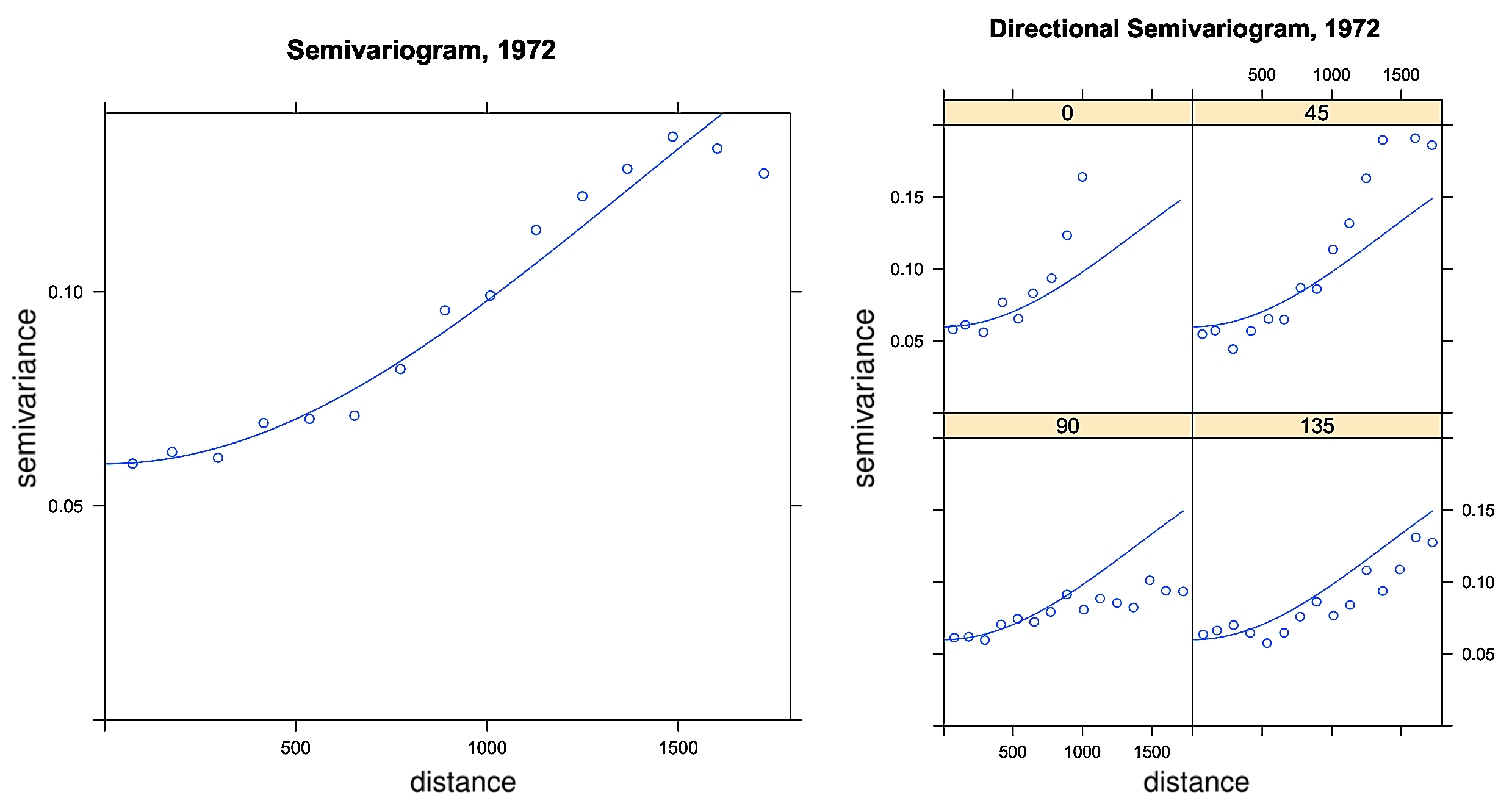}\end{center}

\caption{Semivariogram and directional semivariograms. The data is from the
 year 1972.}
\label{fig:semivariogram1972}
\end{figure}

A potential concern is the isotropy assumption, i.e., the direction
of the vector giving the difference between the locations of two points
in the random field does not matter. That is, for the variogram to
only be a function of the distance between its arguments and not the
direction of the vector from one to the other, we must assume that
the direction does not impact the covariance. One way to assess this
is to plot and inspect the directional variograms, a typical (for
this data) example of which is shown in Figure \ref{fig:semivariogram1972}. 

The directional variograms are reassuring. The empirical directional
variograms seem to align well with the fitted isotropic variogram
up to at least 500 km. Beyond that distance, very little data goes
into estimating each point in the plots; hence they are quite variable.
The appearance of a trend away from the fitted isotropic variogram
can be due to the fact that these points are highly correlated.

We are  interested in fitting the model
\begin{equation}
Y_{i}=\delta_{\mathrm{boy}}1\{i\in\mathcal{B}\}+\delta_{\mathrm{girl}}1\{i\in\mathcal{G}\}+R_{\mathrm{true},i}\left(\beta_{\mathrm{boy}}1\{i\in\mathcal{B}\}+\beta_{\mathrm{girl}}1\{i\in\mathcal{G}\}\right)+F_{i}^{T}\gamma+\epsilon_{i},
\label{eq:maccinimodel}
\end{equation}
where $\mathcal{B}$ and $\mathcal{G}$ are the set of observation
indices corresponding to subjects who are boys and girls, respectively.
The random variables $R_{\mathrm{true},i}$ are sampled from the random
field of the log of the ratio of yearly rainfall to long term average
at the outcome locations, and $F$ includes location (district) dummies,
season dummies, time trend, and interactions. $R_{\mathrm{true},i}$'s
are not observed but $\mathbf{R}^{*}$, a vector of observations from
the same random field albeit at different locations, is observed. 

\begin{figure}
\begin{center}\includegraphics[scale=0.2]{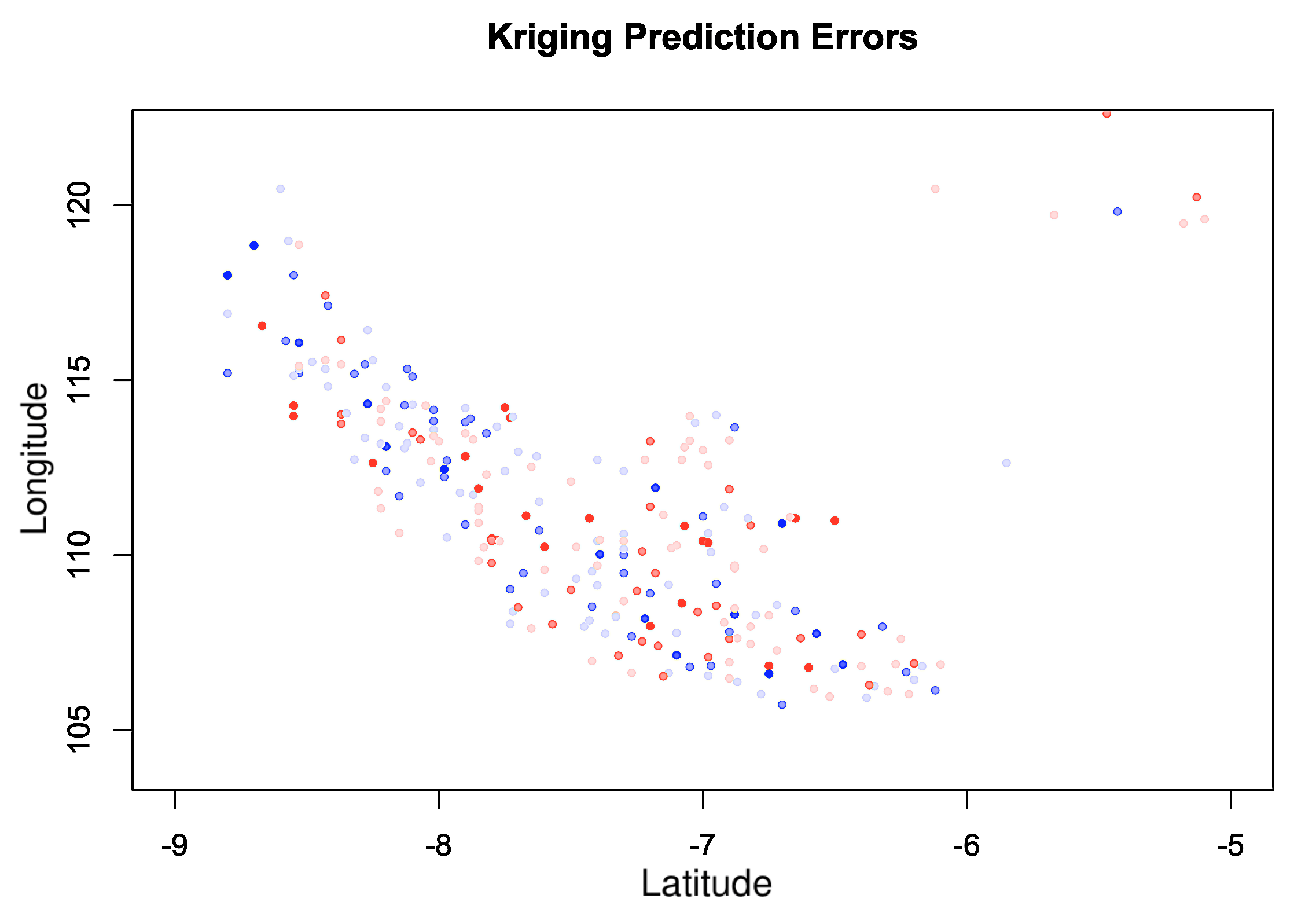}\end{center}

\caption{Geographic plot of prediction errors from leave-one-out cross-validation.
Negative residuals are in blue, positive residuals in red. Darker
shades correspond to greater absolute values.}
\label{fig:looplot}
\end{figure}



Remark that we use a different two-step method than \citet{maccini2009under} did. Whilst we  do the interpolation step with the best linear
predictor, they use an instrumental variables approach. Their strategy
is to estimate \eqref{eq:maccinimodel} by running a two-stage least-squares regression
in which the nearest rainfall is instrumented with slightly more distant
rainfall stations (the second to fifth closest stations). We argue
that this approach is problematic for two reasons. Conceptually, there
is no reason to believe that instrumenting the wrong rainfall measurements
with other wrong rainfall measurements will ``wash away'' the error
term, which can be thought of as the difference between the rainfall
at the outcome location and the rainfall at the nearest station (for
instance, all stations could be near each other and have correlated
``errors''). However, such a strong assumption is necessary for
two-stage least-squares approach to deliver a consistent estimator.
Practically, the data set at hand makes the first-stage regression
coefficient difficult to interpret and makes this strategy tricky
to implement. Indeed, the closer rainfall stations do not have the
same distance or configuration around any outcome location. Furthermore,
for any given outcome location, there will be few years for which
the nearest rainfall station and all of the next four nearest stations
will have observed measurements, so the, say, second nearest station
does not correspond to the same station for every observation.

To further assess the quality of the Kriging estimates, we plot the
geographically located interpolation errors from leave-one-out cross-validation
in Figure \ref{fig:looplot}. The figure suggests that the residuals have low spatial
correlation. 

The second stage for the Krig-and-regress is done by ordinary least-squares.
We find that using feasible generalized least-squares did not help
in this case.

\begin{table}

\centering
\renewcommand{\arraystretch}{1.5}
\begin{tabular}{cccccc}
\toprule
 &  & M\&Y & 2-step (K-R) & 2-step (BS) & Min Dist\\
\midrule
\multirow{2}{*}[-4pt]{(a)} & $\beta_{\mathrm{girl}}$ & $\underset{(0.046)}{0.011}$ & $\underset{(0.055)}{0.126}$ & $\underset{(0.058)}{0.136}$ & $\underset{(0.042)}{0.085}$\\
 & $\beta_{\mathrm{boy}}$ & $\underset{(0.050)}{0.037}$ & $\underset{(0.061)}{0.037}$ & $\underset{(0.058)}{0.049}$ & $\underset{(0.180)}{-0.092}$\\
\cmidrule{2-6}
\multirow{2}{*}[-4pt]{(b)} & $\beta_{\mathrm{girl}}$ & $\underset{(0.055)}{\text{\textminus}0.133}$ & $\underset{(0.065)}{\text{\textminus}0.149}$ & $\underset{(0.078)}{\text{\textminus}0.141}$ & $\underset{(0.212)}{\text{\textminus}0.206}$\\
 & $\beta_{\mathrm{boy}}$ & $\underset{(0.050)}{\text{\textminus}0.044}$ & $\underset{(0.061)}{0.010}$ & $\underset{(0.073)}{0.023}$ & $\underset{(0.127)}{\text{\textminus}0.170}$\\
\cmidrule{2-6}
\multirow{2}{*}[-4pt]{(c)} & $\beta_{\mathrm{girl}}$ & $\underset{(0.896)}{1.631}$ & $\underset{(1.068)}{2.371}$ & $\underset{(1.183)}{2.064}$ & $\underset{(2.012)}{1.613}$\\
 & $\beta_{\mathrm{boy}}$ & $\underset{(1.026)}{0.973}$ & $\underset{(1.263)}{0.599}$ & $\underset{(1.300)}{0.443}$ & $\underset{(2.372)}{0.905}$\\
\cmidrule{2-6}
\multirow{2}{*}[-4pt]{(d)} & $\beta_{\mathrm{girl}}$ & $\underset{(0.529)}{\text{\textminus}0.755}$ & $\underset{(0.630)}{\text{\textminus}1.381}$ & $\underset{(0.580)}{\text{\textminus}1.395}$ & $\underset{(0.445)}{\text{\textminus}1.795}$\\
 & $\beta_{\mathrm{boy}}$ & $\underset{(0.543)}{0.353}$ & $\underset{(0.669)}{0.465}$ & $\underset{(0.768)}{0.440}$ & $\underset{(0.360)}{\text{\textminus}1.431}$\\
\cmidrule{2-6}
\multirow{2}{*}[-4pt]{(e)} & $\beta_{\mathrm{girl}}$ & $\underset{(0.612)}{1.518}$ & $\underset{(0.730)}{1.337}$ & $\underset{(0.817)}{1.346}$ & $\underset{(0.232)}{0.530}$\\
 & $\beta_{\mathrm{boy}}$ & $\underset{(0.679)}{\text{\textminus}0.485}$ & $\underset{(0.835)}{\text{\textminus}1.569}$ & $\underset{(1.100)}{\text{\textminus}1.493}$ & $\underset{(0.212)}{\text{\textminus}2.088}$\\
 \bottomrule
\end{tabular}

\caption{Effect of birth year rainfall on (a) indicator for very good self-reported
health status; (b) indicator for poor or very poor self-reported health
status; (c) adult height; (d) days absent due to illness (during the
four weeks preceding the survey); (e) completed grades of schooling.
The (M\&Y) estimates are computed using the instrumental variables
approach of \citet{maccini2009under}; Krig-and-regress (K-R) uses OLS
in the second stage and the standard errors of the OLS output; the
procedure for the two-step bootstrap (BS) is as detailed in Section
\ref{sssec:bootstrap}, with additional covariates are added in the second stage. }
\label{tab:effect}
\end{table}

The difference between the IV and the Krig-and-regress methods is
noticeable. This should not be surprising as they correspond to completely
different approaches to imputing the missing covariate in the first
step. The economic magnitude of the difference is relevant for policy
implications; the Krig-and-regress point estimate is 0.115 higher
than that of the IV approach. This means that the estimated impact
of the first standard deviation in yearly rainfall, away from the
long term average, on the estimated probability of declaring oneself
very healthy increases from 0.4\% to 3.5\%.

The valid standard errors of the two-step bootstrap tend to be larger
than the naive, unprincipled ones but moderately so and do not seem
to pose a threat to statistical significance in this application;
note that there is nothing inherently conservative about their design.

The only important disparity in the minimum-distance regression output
is for completed grades of schooling. It suggests a significant effect
where Krig-and-regress does not, but for a smaller point estimate
than the IV estimator does in \citet{maccini2009under}. We find that, for
girls, a one standard deviation increase in rainfall brings about
an increase of 0.15 in years of schooling, as opposed to 0.42.

As remarked in the simulation of Section \ref{sec:madsen}, the standard errors of
the minimum-distance estimator can be tangibly different from those
of the Krig-and-regress estimator, sometimes smaller and sometimes
larger, but they were observed to be generally accurate.

\section{Discussion and Conclusion}\label{sec:conclusion}

We developed methods handling regression analysis with misaligned
data and paid particular attention to the case in which the researcher
does not want to specify the covariance structure of the regression
errors, thus making maximum likelihood estimation for the full model
inaccessible. First, we suggested a modified two-step method which
produces confidence intervals that account for the uncertainty due
to the estimation of the imputed regressor and is very easy to compute.
This contribution is important because, although Krig-and-regress
was already recommended for two-step estimation in the misaligned
data regression problem, to the best of our knowledge, the literature
did not offer standard errors that account for estimation in the
first stage without requiring specification of the regression error
covariance structure. Second, we proposed a one-step minimum-distance
estimator and developed its limit distribution theory. We produced
an asymptotic approximation formula for the covariance of the estimator
as well as a likelihood-free sampling strategy that provides pivotal
inference with respect to the covariance matrix of the regression
errors.

\subsection*{Acknowledgements}

First and foremost, I would like to thank my PhD advisors Gary Chamberlain,
Edward Glaeser, Neil Shephard, and Elie Tamer for their guidance and
support. I am indebted for their insightful comments to Alberto Abadie,
Nathaniel Hendren, James Heckman, Guido Imbens, Bruce Meyer, Jann
Spiess, Michael Stein, and Bryce Millett Steinberg. Zhen Xie has provided
outstanding research assistance.

\bibliography{references}

\begin{thebibliography}{32}
\providecommand{\natexlab}[1]{#1}
\providecommand{\url}[1]{\texttt{#1}}
\expandafter\ifx\csname urlstyle\endcsname\relax
  \providecommand{\doi}[1]{doi: #1}\else
  \providecommand{\doi}{doi: \begingroup \urlstyle{rm}\Url}\fi

\bibitem[Abramowitz and Stegun(1964)]{abramowitz1964handbook}
Abramowitz, Milton and Stegun, Irene~A.
\newblock \emph{Handbook of mathematical functions}, volume 1046.
\newblock Dover, New York, 1964.

\bibitem[Chay and Greenstone(1999)]{chay2003impact}
Chay, Kenneth~Y and Greenstone, Michael.
\newblock The impact of air pollution on infant mortality: Evidence from
  geographic variation in pollution shocks induced by a recession.
\newblock Working Paper 7442, National Bureau of Economic Research, December
  1999.
\newblock URL \url{http://www.nber.org/papers/w7442}.

\bibitem[Chernozhukov and Hong(2003)]{chernozhukov2003mcmc}
Chernozhukov, Victor and Hong, Han.
\newblock An mcmc approach to classical estimation.
\newblock \emph{Journal of Econometrics}, 115\penalty0 (2):\penalty0 293--346,
  2003.

\bibitem[Cressie(2015)]{cressie2015statistics}
Cressie, Noel.
\newblock \emph{Statistics for spatial data}.
\newblock John Wiley \& Sons, 2015.

\bibitem[Dell et~al.(2014)Dell, Jones, and Olken]{dell2014we}
Dell, Melissa, Jones, Benjamin~F, and Olken, Benjamin~A.
\newblock What do we learn from the weather? the new climate-economy
  literature.
\newblock \emph{Journal of Economic Literature}, 52\penalty0 (3):\penalty0
  740--98, 2014.

\bibitem[Diggle et~al.(2007)Diggle, Ribeiro, and
  Geostatistics]{diggle2007springer}
Diggle, PJ, Ribeiro, PJ, and Geostatistics, Model-based.
\newblock Springer series in statistics, 2007.

\bibitem[Fabregas et~al.(2017)Fabregas, Kremer, Robinson, and
  Schilbach]{fabregas2017institutions}
Fabregas, Raissa, Kremer, Michael, Robinson, Jon, and Schilbach, Frank.
\newblock What institutions are appropriate for generating and disseminating
  local agricultural information?, 2017.

\bibitem[Forneron and Ng(2018)]{forneron2018abc}
Forneron, Jean-Jacques and Ng, Serena.
\newblock The abc of simulation estimation with auxiliary statistics.
\newblock \emph{Journal of Econometrics}, 205\penalty0 (1):\penalty0 112--139,
  2018.

\bibitem[Gelfand et~al.(2010)Gelfand, Diggle, Guttorp, and
  Fuentes]{gelfand2010handbook}
Gelfand, Alan~E, Diggle, Peter, Guttorp, Peter, and Fuentes, Montserrat.
\newblock \emph{Handbook of spatial statistics}.
\newblock CRC press, 2010.

\bibitem[Gelman et~al.(2013)Gelman, Carlin, Stern, Dunson, Vehtari, and
  Rubin]{gelman2013bayesian}
Gelman, Andrew, Carlin, John~B, Stern, Hal~S, Dunson, David~B, Vehtari, Aki,
  and Rubin, Donald~B.
\newblock \emph{Bayesian data analysis}.
\newblock CRC press, 2013.

\bibitem[Herlihy et~al.(1998)Herlihy, Stoddard, and
  Johnson]{herlihy1998relationship}
Herlihy, Alan~T, Stoddard, John~L, and Johnson, Colleen~Burch.
\newblock The relationship between stream chemistry and watershed land cover
  data in the mid-atlantic region, us.
\newblock \emph{Water, Air, and Soil Pollution}, 105\penalty0 (1-2):\penalty0
  377--386, 1998.

\bibitem[Hoff(2009)]{hoff2009first}
Hoff, Peter~D.
\newblock \emph{A first course in Bayesian statistical methods}, volume 580.
\newblock Springer, 2009.

\bibitem[Jenish and Prucha(2009)]{jenish2009central}
Jenish, Nazgul and Prucha, Ingmar~R.
\newblock Central limit theorems and uniform laws of large numbers for arrays
  of random fields.
\newblock \emph{Journal of econometrics}, 150\penalty0 (1):\penalty0 86--98,
  2009.

\bibitem[Jenish and Prucha(2012)]{jenish2012spatial}
Jenish, Nazgul and Prucha, Ingmar~R.
\newblock On spatial processes and asymptotic inference under near-epoch
  dependence.
\newblock \emph{Journal of econometrics}, 170\penalty0 (1):\penalty0 178--190,
  2012.

\bibitem[Jiang(1997)]{jiang1997derivation}
Jiang, Jiming.
\newblock A derivation of blup-best linear unbiased predictor.
\newblock \emph{Statistics \& Probability Letters}, 32\penalty0 (3):\penalty0
  321--324, 1997.

\bibitem[Lahiri(2003)]{lahiri2003central}
Lahiri, SN.
\newblock Central limit theorems for weighted sums of a spatial process under a
  class of stochastic and fixed designs.
\newblock \emph{Sankhy{\=a}: The Indian Journal of Statistics}, pages 356--388,
  2003.

\bibitem[Lahiri et~al.(2002)Lahiri, Lee, and Cressie]{lahiri2002asymptotic}
Lahiri, Soumendra~Nath, Lee, Yoondong, and Cressie, Noel.
\newblock On asymptotic distribution and asymptotic efficiency of least squares
  estimators of spatial variogram parameters.
\newblock \emph{Journal of Statistical Planning and Inference}, 103\penalty0
  (1-2):\penalty0 65--85, 2002.

\bibitem[Maccini and Yang(2009)]{maccini2009under}
Maccini, Sharon and Yang, Dean.
\newblock Under the weather: Health, schooling, and economic consequences of
  early-life rainfall.
\newblock \emph{American Economic Review}, 99\penalty0 (3):\penalty0 1006--26,
  2009.

\bibitem[Madsen et~al.(2008)Madsen, Ruppert, and Altman]{madsen2008regression}
Madsen, Lisa, Ruppert, David, and Altman, Naomi~S.
\newblock Regression with spatially misaligned data.
\newblock \emph{Environmetrics: The official journal of the International
  Environmetrics Society}, 19\penalty0 (5):\penalty0 453--467, 2008.

\bibitem[Matheron(1962)]{matheron1962traite}
Matheron, Georges.
\newblock \emph{Trait{\'e} de g{\'e}ostatistique appliqu{\'e}e.}
\newblock Editions Technip, 1962.

\bibitem[Miguel et~al.(2004)Miguel, Satyanath, and
  Sergenti]{miguel2004economic}
Miguel, Edward, Satyanath, Shanker, and Sergenti, Ernest.
\newblock Economic shocks and civil conflict: An instrumental variables
  approach.
\newblock \emph{Journal of political Economy}, 112\penalty0 (4):\penalty0
  725--753, 2004.

\bibitem[Murphy and Topel(1985)]{murphy1985estimation}
Murphy, Kevin~M and Topel, Robert~H.
\newblock {Estimation and Inference in Two-Step Econometric Models}.
\newblock \emph{Journal of Business \& Economic Statistics}, 3\penalty0
  (4):\penalty0 370--379, October 1985.
\newblock URL \url{https://ideas.repec.org/a/bes/jnlbes/v3y1985i4p370-79.html}.

\bibitem[Nunn and Puga(2012)]{nunn2012ruggedness}
Nunn, Nathan and Puga, Diego.
\newblock Ruggedness: The blessing of bad geography in africa.
\newblock \emph{Review of Economics and Statistics}, 94\penalty0 (1):\penalty0
  20--36, 2012.

\bibitem[Pagan(1984)]{pagan1984econometric}
Pagan, Adrian.
\newblock Econometric issues in the analysis of regressions with generated
  regressors.
\newblock \emph{International Economic Review}, 25\penalty0 (1):\penalty0
  221--247, 1984.
\newblock ISSN 00206598, 14682354.
\newblock URL \url{http://www.jstor.org/stable/2648877}.

\bibitem[Phillips et~al.(1992)Phillips, Dolph, and
  Marks]{phillips1992comparison}
Phillips, Donald~L, Dolph, Jayne, and Marks, Danny.
\newblock A comparison of geostatistical procedures for spatial analysis of
  precipitation in mountainous terrain.
\newblock \emph{Agricultural and forest meteorology}, 58\penalty0
  (1-2):\penalty0 119--141, 1992.

\bibitem[Pouliot(2016)]{pouliot2016missing}
Pouliot, Guillaume.
\newblock \emph{Missing Data Problems}.
\newblock PhD thesis, Harvard University, 2016.

\bibitem[Pouliot(2020)]{pouliot2020lecture}
Pouliot, Guillaume~A.
\newblock Lecture notes: Optimization-conscious econometrics, 2020.
\newblock URL \url{https://sites.google.com/site/guillaumeallairepouliot/}.

\bibitem[Qu et~al.(2017)Qu, Lee, and Yu]{qu2017qml}
Qu, Xi, Lee, Lung-fei, and Yu, Jihai.
\newblock Qml estimation of spatial dynamic panel data models with endogenous
  time varying spatial weights matrices.
\newblock \emph{Journal of Econometrics}, 197\penalty0 (2):\penalty0 173--201,
  2017.

\bibitem[Shah and Steinberg(2013)]{shah2013drought}
Shah, Manisha and Steinberg, Bryce~Millett.
\newblock Drought of opportunities: Contemporaneous and long term impacts of
  rainfall shocks on human capital.
\newblock Working Paper 19140, National Bureau of Economic Research, June 2013.
\newblock URL \url{http://www.nber.org/papers/w19140}.

\bibitem[Stein(2012)]{stein2012interpolation}
Stein, Michael~L.
\newblock \emph{Interpolation of spatial data: some theory for kriging}.
\newblock Springer Science \& Business Media, 2012.

\bibitem[Tabios~III and Salas(1985)]{tabios1985comparative}
Tabios~III, Guillermo~Q and Salas, Jose~D.
\newblock A comparative analysis of techniques for spatial interpolation of
  precipitation 1.
\newblock \emph{JAWRA Journal of the American Water Resources Association},
  21\penalty0 (3):\penalty0 365--380, 1985.

\bibitem[Xu and Lee(2015)]{xu2015maximum}
Xu, Xingbai and Lee, Lung-fei.
\newblock Maximum likelihood estimation of a spatial autoregressive tobit
  model.
\newblock \emph{Journal of Econometrics}, 188\penalty0 (1):\penalty0 264--280,
  2015.

\end{thebibliography}

\pagebreak{}

\appendix
\appendixpage

\section{Covariance Matrix Estimation}

\subsection*{Computing the Covariance of the Minimum-Distance Estimator}

Some comments are in order regarding the computation of $\Sigma_{g}(\theta_{0})$,
in particular the fourth-order terms involved in the computation of
\emph{$\sigma_{12}(x)=Cov_{\theta_{0}}\left(\left(\varepsilon(0)-\varepsilon(d_{1})\right)^{2},\left(\varepsilon(x)-\varepsilon(x+d_{2})\right)^{2}\right)$}
for $d_{1},d_{2}\in\mathbb{R}^{2}$. This may be expressed as
\[
E\left[\left(\varepsilon(0)-\varepsilon(d_{1})\right)^{2}\left(\varepsilon(x)-\varepsilon(x+d_{2})\right)^{2}\right] - E\left[\left(\varepsilon(0)-\varepsilon(d_{1})\right)^{2}\right]E\left[\left(\varepsilon(x)-\varepsilon(x+d_{2})\right)^{2}\right],
\]
and we can write 
\[
E\left[\left(\varepsilon(0)-\varepsilon(d_{1})\right)^{2}\left(\varepsilon(x)-\varepsilon(x+d_{2})\right)^{2}\right]
\]
\[
=\int_{|x|}\int_{\angle x}E_{\varepsilon}\left[\left(\varepsilon(0)-\varepsilon(d_{1})\right)^{2}\left(\varepsilon(x)-\varepsilon(x+d_{2})\right)^{2}\right]f_{x}(|x|,\angle x)d|x|d\angle x,
\]
where $|x|$ and $\angle x$ are the length and angle, respectively,
of a location $x\in\mathbb{R}^{2}$ considered as a vector from the
origin. The first order of business is to approximate the inner expectation.
This is straightforward since 
\begin{align*}
&\phantom{=}
E_{\varepsilon}\left[\left(\varepsilon(0)-\varepsilon(d_{1})\right)^{2}\left(\varepsilon(x)-\varepsilon(x+d_{2})\right)^{2}\right]
\\
&=
E_{\varepsilon}\left[\left(\varepsilon(0)^{2}-2\varepsilon(0)\varepsilon(d_{1})+\varepsilon(d_{1})^{2}\right)\left(\varepsilon(x)^{2}-2\varepsilon(x)\varepsilon(x+d_{2})+\varepsilon(x+d_{2})^{2}\right)\right]
\\
&=
E_{\varepsilon}[\varepsilon(0)^{2}\varepsilon(x)^{2}-2\varepsilon(0)^{2}\varepsilon(x)\varepsilon(x+d_{2})+\varepsilon(0)^{2}\varepsilon(x+d_{2})^{2}
\\
&\phantom{=} \quad
- 2\varepsilon(0)\varepsilon(d_{1})\varepsilon(x)^{2}+4\varepsilon(0)\varepsilon(d_{1})\varepsilon(x)\varepsilon(x+d_{2})-2\varepsilon(0)\varepsilon(d_{1})\varepsilon(x+d_{2})^{2}
\\
&\phantom{=} \quad
+ \varepsilon(d_{1})^{2}\varepsilon(x)^{2}-2\varepsilon(d_{1})^{2}\varepsilon(x)\varepsilon(x+d_{2})+\varepsilon(d_{1})^{2}\varepsilon(x+d_{2})^{2}],
\end{align*}
where all fourth-order moments may be approximated using Isserlis'
formula, which is in terms of second-order moments, for which we have
reliable nonparametric estimates.

\subsection*{Covariance Estimation for the Minimum-Distance Estimator with Isotropic
Variogram}

Theorem \ref{thm:1} gives a limit distribution for vectors of anisotropic variograms,
which take as arguments the distance and angle between the two points
whose covariance we evaluate (as opposed to isotropic variograms,
which take only the distance). The reason is that, without making
cavalier assumptions about $f$, the anisotropic variogram is not
weakly stationary (even when the random field of its arguments is
strongly stationary).

However, one can easily use Theorem \ref{thm:1} to obtain the limit distribution
of a minimum-distance estimator under isotropy assumption. Suppose
there are $p=p(n)$ bins for directions (e.g., the default binning
of the statistical package used) with direction bin centers $\varrho_{1},...,\varrho_{p}$.
Denote vectors of length $r$ and angle $\varrho$ by $(r,\varrho)$.
Slightly abusing notation, also denote the anisotropic variogram by
$\gamma_{YR^{*}}$ and $\gamma_{R^{*}}$, distinguishing them by whether
their argument is a distance scalar or a vector, and estimate the
statistic
\[
g_{n}^{\mathrm{iso}}(\phi)=2\left(\gamma_{YR^{*}}^{*}(r_{1},\varrho_{1})-\gamma_{YR^{*}}(r_{1};\phi),...,\gamma_{YR^{*}}^{*}(r_{1},\varrho_{p})-\gamma_{YR^{*}}(r_{1};\phi),\right.
\]
\[
\left.\gamma_{R^{*}}^{*}(r_{2},\varrho_{1})-\gamma_{R^{*}}(r_{2};\phi),...,\gamma_{R^{*}}^{*}(r_{2},\varrho_{p})-\gamma_{YR}(r_{2};\phi)\right).
\]
Note that we enforce the covariance parameter $\phi$ to be the
same regardless of the direction $\varrho_{i}$, $i=1,...,p$. With
this ``trick'', we can leverage the efficiency gain from the isotropy
assumption and obtain standard errors for the resulting minimum-distance
estimator using Theorem \ref{thm:1}. 

\pagebreak{}

\section{Distribution Theory for Minimum-Distance Estimation}\label{app:distribution}

Limit distribution theory depends on the asymptotic domain chosen
by the analyst. One may consider the pure-increasing domain (more
and more points, always at more than some minimum distance from each
other), infill asymptotics (more and more points in a fixed, finite
area), or a mix of the two (points get denser and extend over a wider
area as their number increases). There is a deep conceptual difference
between infill asymptotics and increasing domain asymptotics, pure
or mixed. For instance, the infill domain framework may not allow
for consistent estimation of mean parameters (see Lahiri, 1996). The
difference between mixed- and pure-increasing domain asymptotics is,
on the other hand, of rather technical nature. For instance, we find
that equivalent results are obtained for the limit distribution theory
of the directional variograms in both the pure- and mixed-increasing
domains. To be sure, both yield the same asymptotic variance approximation.
The area under study will be large enough (compared to the range of
the spatial correlations) that the natural choice is to use the increasing
domain framework in our main application.

In order to obtain a central limit theorem for the minimum-distance
estimator, we must first obtain a central limit theorem for the statistics
from which we want to minimize distance. The following lemma is a
useful preliminary result.

Let $P_{X}$ denote the joint probability distribution of the sequence
of iid random location vectors $X_{1},X_{2},...$ with density $f$,
whose realization are denoted $x_{1},x_{2},...$. Recall that the
analysis is conditional on the location vectors.
\begin{lemma}
\label{lem:b1}
Suppose that $\left\{ Z(x):x\in\mathbb{R}^{d}\right\} $
is a stationary random field with $EZ(0)=0$ such that $E\left|Z(0)\right|^{2+\delta}<\infty$
for some $\delta>0$. Suppose $f$ is continuous and everywhere positive
on $\overline{\mathcal{R}}_{0}$, and that $\int_{\mathcal{R}_{0}}f^{2}(x)dx<\infty$.
Let $\alpha_{1}(a)=a^{-\tau}$ for some $\tau>\frac{d(2+\delta)}{\delta}$
and suppose $g(b)=o\left(b^{\frac{\tau-d}{4d}}\right)$. Further suppose
that $\left(\log n\right)^{2}\lambda_{n}^{\frac{d-\tau}{4\tau}}\rightarrow0$
as $n\rightarrow\infty$.
\renewcommand{\labelenumi}{(\roman{enumi})}
\begin{enumerate}
\item If $n/\lambda_{n}^{d}\rightarrow C_{1}\in(0,\infty)$ as
$n\rightarrow\infty$, then 
\[
n^{-\frac{1}{2}}\sum_{i=1}^{n}Z(x_{i})\overset{d}{\rightarrow}N\left(0,\sigma(0)+C_{1}\cdot Q\cdot\int_{\mathbb{R}^{d}}\sigma(x)dx\right),
\]
a.s. $P_{X}$, where $Q=\int_{\mathcal{R}_{0}}f^{2}(x)dx$ and $\sigma(d)=EZ(0)Z(d)$.
\item If $n/\lambda_{n}^{d}\rightarrow\infty$ as $n\rightarrow\infty$,
then 
\[
\frac{\lambda_{n}^{\frac{d}{2}}}{n}\sum_{i=1}^{n}Z(x_{i})\overset{d}{\rightarrow}N\left(Q\cdot\int_{\mathbb{R}^{d}}\sigma(x)dx\right)
\]
a.s. $P_{X}$.
\end{enumerate}
\end{lemma}



\begin{Proof}
The claim follows directly from Proposition 3.1, Theorem 3.1 and Theorem
3.2 of \citet{lahiri2003central}.
\end{Proof}

We prove a more general result than stated in the main body of the
article to allow for a general location dimension $d$ and for covariates
in the deterministic component of $R(x)=s(x)^{T}\rho+\varepsilon(x)$.
Let the mean of the random field be $s(x)^{T}\rho$ , and replace
the assumption $\lambda_{n}^{2}\left\Vert \hat{m}-m\right\Vert _{2}^{4}=o_{p}(1)$
by $\sup\left\{ \left\Vert s(x)-s(x+h)\right\Vert _{2}^{2}:x\in\mathbb{R}^{d}\right\} \le C(h)<\infty$
, $\lambda_{n}^{d}\left\Vert \hat{\rho}-\rho\right\Vert _{2}^{4}=o_{p}(1)$,
and\emph{ }$\left\Vert s(x_{j})-s(x_{i}+h)\right\Vert _{2}=O\left(\lambda_{n}^{-\frac{d}{4}}\right)$
for any \emph{$(i,j)$} pair in any given bin of the nonparametric
variogram.

We will also need the following technical lemma.
\begin{lemma}{(Lemma 5.2, \citealp{lahiri2003central})}
\label{lem:b2}
Let
$\left\{ Z(x):x\in\mathbb{R}^{d}\right\} $ be a stationary random
field with $E\left[Z(0)\right]=0$. Suppose that $\int\left|\sigma(x)\right|dx<\infty$.
Suppose that $f$ is continuous and everywhere positive on $\overline{\mathcal{R}}_{0}$,
and that $\int_{\mathcal{R}_{0}}f^{2}(x)dx<\infty$. Suppose that
there exists a function $Q_{1}(\cdot)$ such that 
\begin{align*}
\left(\int \omega^{2}(\lambda_{n}x)f(x)dx\right)^{-1}\int \omega(\lambda_{n}x)\omega(x'+\lambda_{n}x)f^{2}(x)dx\rightarrow Q_{1}(x'), \quad
\forall x'\in\mathbb{R}^{d}.
\end{align*}
Then, in both pure- and mixed-increasing
domain asymptotics,
\begin{align*}
E\left(\sum_{i=1}^{n}\omega\left(x_{i}\right)Z\left(x_{i}\right)\right)^{2}=O(\lambda_{n}^{-d}n^{2}) \quad as.\ P_X,
\end{align*}
and for any $\epsilon>0$,
\useshortskip
\begin{align*}
\left(\sum_{i=1}^{n}\omega\left(x_{i}\right)Z\left(x_{i}\right)\right)^{2}=o_{p}(\lambda_{n}^{-d+\epsilon}n^{2}) \quad as.\ P_X.
\end{align*}
\end{lemma}

\begin{Proof}
Let $\mathcal{Z}_{n}=\left(\sum_{i=1}^{n}\omega\left(x_{i}\right)Z\left(x_{i}\right)\right)^{2}$.
Lemma 5.2 of Lahiri (2003) guarantees that $E\mathcal{Z}_{n}=O(\lambda_{n}^{-d}n^{2})$.
Specifically, $\exists$ $M$ and $N$ such that $E\mathcal{Z}_{n}/\lambda_{n}^{-d}n^{2}\le M$,
$\forall\ $ $n\ge N$. Consequently, for any $\epsilon > 0$, $E\mathcal{Z}_{n}/\left(\lambda_{n}^{-d+\epsilon}n^{2}\right)\le M/\lambda_{n}^{\epsilon},\forall n\ge N.$
Therefore, given any $\delta>0$, there exists $N'>N$ such that $M/\lambda_{n}^{\epsilon}<\delta$
for all $n>N'$, which implies that $E\mathcal{Z}_{n}/\left(\lambda_{n}^{-d+\epsilon}n^{2}\right)\le\delta$,
$\forall$ $n\ge N'$, meaning it converges to zero.

By Markov's inequality, we have that for any $c>0$, 
\[
P\left(\left|\lambda_{n}^{d-\epsilon}n^{-2}\mathcal{Z}_{n}\right|>c\right)\le\frac{E\left|\lambda_{n}^{d-\epsilon}n^{-2}\mathcal{Z}_{n}\right|}{c},
\]
where the right-hand side has been found to be $o(1)$. Consequently,
$\mathcal{Z}_{n}=o_{p}(\lambda_{n}^{-d+\epsilon}n^{2})$.
\end{Proof}

We are interested in the specific nonparametric variogram 
\[
\hat{\gamma}(h)=\frac{1}{\left|N_{n}(h)\right|}\sum_{(i,j)\in N_{n}(h)}\left(\hat{\varepsilon}(x_{i})-\hat{\varepsilon}(x_{j})\right)^{2},
\]
where $\hat{\varepsilon}(x)=R(x)-s(x)^{T}\hat{\rho}$, and $\hat{\rho}$
is an estimate of $\rho$, typically a least-squares regression
coefficient estimate. We want to characterize the limiting behavior
of $g_{n}(\phi)=\left(\hat{\gamma}(h_{k})-\gamma(h_{k};\phi)\right)_{k=1}^{K}$.

\renewcommand{\thetheorem}{1}
\begin{theorem}
Suppose that $\left\{ \varepsilon(x):x\in\mathbb{R}^{d}\right\} $
is a stationary random field
such that $E\left|\varepsilon(0)\right|^{4+\delta}<\infty$
for some $\delta>0$. Suppose $f$ is continuous and everywhere positive
on $\overline{\mathcal{R}}_{0}$, and that $\int_{\mathcal{R}_{0}}f^{2}(x)dx<\infty$.
Let $\alpha_{1}(a)=a^{-\tau}$ for some $\tau>\frac{d(2+\delta)}{\delta}$
and suppose $g(b)=o\left(b^{\frac{\tau-d}{4d}}\right)$. Suppose that
$\left(\log n\right)^{2}\lambda_{n}^{\frac{d-\tau}{4\tau}}\rightarrow0$
as $n\rightarrow\infty$. Further suppose that the autocovariance function
$\sigma_{ij}(x)=Cov_{\phi_{0}}\left(\left(\varepsilon(0)-\varepsilon(h_{i})\right)^{2},\left(\varepsilon(x)-\varepsilon(x+h_{j})\right)^{2}\right)$
satisfies $\int\left|\sigma_{ij}(x)\right|dx<\infty$, $i,j=1,...,K$.
Suppose that $\lambda_{n}^{d}\left\Vert \hat{\rho}-\rho\right\Vert _{2}^{4}=o_{p}(1)$,
$\left|N_{n}(h_{k})\right|=\left(1+o(1)\right)n$, $\left|N_{r,n}(h_{k})\right|=o(\lambda_{n}^{d})$,
$\left|N'_{n}(h_{k})\right|=o(\lambda_{n}^{d})$, and
$E\left[\left|\left(\varepsilon(x_{i})-\varepsilon(x_{j})\right)^{2}-\left(\varepsilon(x_{i})-\varepsilon(x_{i}+h_{k})\right)^{2}\right|\right]=o(\lambda_{n}^{-\frac{d}{2}})$
for
all $i,j\in N_{n}(h_{k})$, $k=1,...,K$. Suppose that $\sup\left\{ \left\Vert s(x)-s(x+h)\right\Vert _{2}^{2}:x\in\mathbb{R}^{d}\right\} \le C(h)<\infty$
for all $h\in\mathbb{R}^{d}$, and that
$\left\Vert s(x_{j})-s(x_{i}+h_{k})\right\Vert _{2}=O\left(\lambda_{n}^{-\frac{d}{4}}\right)$
for $(i,j)\in N_{n}(h_{k})$ for any $k=1,...,K$. Suppose that
there exists a function $Q_{1}(\cdot)$ such that $\left(\int s_{j}^{2}(\lambda_{n}x)f(x)dx\right)^{-1}\int s_{j}(\lambda_{n}x)s_{j}(x'+\lambda_{n}x)f^{2}(x)dx\rightarrow Q_{1}(x')$, for $j=1,...,p,$
$\forall$ $x'\in\mathbb{R}^{d}$.
\renewcommand{\labelenumi}{(\roman{enumi})}
\begin{enumerate}
\item If $n/\lambda_{n}^{d}\rightarrow C_{1}\in(0,\infty)$ as
$n\rightarrow\infty$, then
\[
n^{\frac{1}{2}}g_{n}(\phi_{0})\overset{d}{\rightarrow}N\left(0,\Sigma_{g}(\phi_{0})\right) \quad a.s.\ P_{X},
\]
where the $i,j$ entry of the covariance matrix is
$\left(\Sigma_{g}(\phi_{0})\right)_{ij}=\sigma_{ij}(0)+Q\cdot C_{1}\cdot\int_{\mathbb{R}^{d}}\sigma_{ij}(x)dx$,
with $Q=\int_{\mathcal{R}_{0}}f^{2}(x)dx$.
\item If $n/\lambda_{n}^{d}\rightarrow\infty$ as $n\rightarrow\infty$,
then
\[
\lambda_{n}^{\frac{d}{2}}g_{n}(\phi_{0})\overset{d}{\rightarrow}N\left(0,\Sigma_{g}(\phi_{0})\right) \quad a.s.\ P_{X},
\]
where $\left(\Sigma_{g}(\theta_{0})\right)_{ij}=Q\cdot\int_{\mathbb{R}^{2}}\sigma_{ij}(x)dx$.
\end{enumerate}
\end{theorem}

\begin{Proof}

Consider the pure-increasing domain. Using the Cram\'er-Wold device,
it suffices to show that $n^{\frac{1}{2}}a^{T}g_{n}(\phi_{0})\overset{d}{\rightarrow}N\left(0,a^{T}\Sigma_{g}(\phi_{0})a\right)$
as $n\rightarrow\infty$ for any $a\in\mathbb{R}^{k}$. Let $g_{0n}=n^{\frac{1}{2}}a^{T}g_{n}(\phi_{0})$,
\[
g_{1n}=n^{1/2}\sum_{k=1}^{K}a_{k}\left(\frac{1}{\left|N_{n}(h_{k})\right|}\sum_{N_{n}(h_{k})}\left(\varepsilon(x_{i})-\varepsilon(x_{j})\right)^{2}-\gamma(h_{k};\phi_{0})\right),
\]
and 
\[
g_{2n}=n^{1/2}\sum_{k=1}^{K}a_{k}\left(\frac{1}{n}\sum_{i=1}^{n}\left(\varepsilon(x_{i})-\varepsilon(x_{i}+h_{k})\right)^{2}-\gamma(h_{k};\phi_{0})\right).
\]

The strategy is to show that, up to an $o_{p}(1)$ difference, $g_{n}$
is close to $g_{1n}$, which is close to $g_{2n}$, which satisfies
the conditions of the central limit theorem obtained in Lemma B.1.
Note that
\[
\left|g_{1n}-g_{0n}\right|\le n^{1/2}\sum_{k=1}^{K}\frac{|a_{k}|}{\left|N_{n}(h_{k})\right|}\left|\sum_{N_{n}(h_{k})}\left(\left(\hat{\varepsilon}(x_{i})-\hat{\varepsilon}(x_{j})\right)^{2}-\left(\varepsilon(x_{i})-\varepsilon(x_{j})\right)^{2}\right)\right|
\]
\[
=n^{1/2}\sum_{k=1}^{K}\frac{|a_{k}|}{\left|N_{n}(h_{k})\right|}\left|\sum_{N_{n}(h_{k})}\left(\left(\left(\hat{\rho}-\rho\right)^{T}\left(s(x_{i})-s(x_{j})\right)-\left(\varepsilon(x_{i})-\varepsilon(x_{j})\right)\right)^{2}-\left(\varepsilon(x_{i})-\varepsilon(x_{j})\right)^{2}\right)\right|
\]
\[
\le n^{1/2}\sum_{k=1}^{K}\frac{|a_{k}|}{\left|N_{n}(h_{k})\right|}\left(\left|\sum_{N_{n}(h_{k})}\left(\left(\hat{\rho}-\rho\right)^{T}\left(s(x_{i})-s(x_{j})\right)\right)^{2}\right|+\left|2\sum_{N_{n}(h_{k})}\left(\hat{\rho}-\rho\right)^{T}\left(s(x_{i})-s(x_{j})\right)\left(\varepsilon(x_{i})-\varepsilon(x_{j})\right)\right|\right).
\]

For the first term, simply observe that by triangle inequality and
the Cauchy-Schwartz inequality,
\begin{align*}
&\phantom{=}
\frac{1}{\left|N_{n}(h_{k})\right|}\left|\sum_{N_{n}(h_{k})}\left(\left(\hat{\rho}-\rho\right)^{T}\left(s(x_{i})-s(x_{j})\right)\right)^{2}\right| \\
&\le\frac{1}{\left|N_{n}(h_{k})\right|}\sum_{N_{n}(h_{k})}\left\Vert \hat{\rho}-\rho\right\Vert _{2}^{2}\left\Vert s(x_{i})-s(x_{j})\right\Vert _{2}^{2} \\
&=o_{p}\left(\lambda_{n}^{-\frac{d}{2}}\right),\numberthis\label{eq:rhos}
\end{align*}
since $\left\Vert \hat{\rho}-\rho\right\Vert _{2}^{2}=o_{p}\left(\lambda_{n}^{-\frac{d}{2}}\right)$.

For the second term, consider the shorthands $\Gamma_{i}^{\varepsilon}=\varepsilon(x_{i})-\varepsilon(x_{i}+h_{k})$
for the individual exact differences on the random fields, and $\triangle_{i,j}^{\varepsilon}=\varepsilon(x_{i}+h_{k})-\varepsilon(x_{j})$
for the approximation error on the random field. Define the analogous
quantities for the covariates, $\Gamma_{i}^{s}=s(x_{i})-s(x_{i}+h_{k})$
and $\triangle_{i,j}^{s}=s(x_{i}+h_{k})-s(x_{j})$. Note that the
dependence on $k$ is implicit. Observe that
\begin{align*}
&\phantom{=}
\frac{1}{\left|N_{n}(h_{k})\right|}\left|\sum_{N_{n}(h_{k})}\left(\hat{\rho}-\rho\right)^{T}\left(s(x_{i})-s(x_{j})\right)\left(\varepsilon(x_{i})-\varepsilon(x_{j})\right)\right| \\
&=
\frac{1}{\left|N_{n}(h_{k})\right|}\left|\sum_{N_{n}(h_{k})}\left(\hat{\rho}-\rho\right)^{T}\left(\Gamma_{i}^{s}+\triangle_{i,j}^{s}\right)\left(\Gamma_{i}^{\varepsilon}+\triangle_{i,j}^{\varepsilon}\right)\right| 
\\
&=
\frac{1}{\left|N_{n}(h_{k})\right|}\left\Vert \hat{\rho}-\rho\right\Vert _{2}\left(\left\Vert \sum_{N_{n}(h_{k})}\Gamma_{i}^{s}\Gamma_{i}^{\varepsilon}\right\Vert _{2}+\left\Vert \sum_{N_{n}(h_{k})}\triangle_{i,j}^{s}\Gamma_{i}^{\varepsilon}\right\Vert _{2}+\left\Vert \sum_{N_{n}(h_{k})}\left(\Gamma_{i}^{s}+\triangle_{i,j}^{s}\right)\triangle_{i,j}^{\varepsilon}\right\Vert _{2}\right).
\numberthis
\label{eq:gamma delta}
\end{align*}

For the first summand of (\ref{eq:gamma delta}), note that
\begin{align*}
&\phantom{=}
\frac{1}{\left|N_{n}(h_{k})\right|}\left\Vert \sum_{N_{n}(h_{k})}\Gamma_{i}^{s}\Gamma_{i}^{\varepsilon}\right\Vert _{2}
\\
&=\frac{1}{\left|N_{n}(h_{k})\right|}\sqrt{\sum_{b=1}^{p}\left(\sum_{N_{n}(h_{k})}\Gamma_{i,b}^{s}\Gamma_{i}^{\varepsilon}\right)^{2}}
\\
&=\frac{1}{\left|N_{n}(h_{k})\right|}\sqrt{\sum_{b=1}^{p}o_{p}(\lambda_{n}^{-\frac{d}{2}}n^{2})}=o_{p}(\lambda_{n}^{-\frac{d}{4}}),
\numberthis \label{eq:summand1}
\end{align*}
by applying Lemma \ref{lem:b2} for $\omega(x_{i})=\Gamma_{i,b}^{s}$ and $\epsilon=\frac{d}{2}$.

For the second summand of (\ref{eq:gamma delta}), note that
\[
\frac{1}{\left|N_{n}(h_{k})\right|}\left\Vert \sum_{N_{n}(h_{k})}\triangle_{i,j}^{s}\Gamma_{i}^{\varepsilon}\right\Vert _{2}\le\frac{1}{\left|N_{n}(h_{k})\right|}\sum_{N_{n}(h_{k})}\left\Vert \triangle_{i,j}^{s}\right\Vert _{2}\left\Vert \Gamma_{i}^{\varepsilon}\right\Vert _{2}=\left|O\left(\lambda_{n}^{-\frac{d}{4}}\right)\right|\frac{1}{\left|N_{n}(h_{k})\right|}\sum_{N_{n}(h_{k})}\left\Vert \Gamma_{i}^{\varepsilon}\right\Vert _{2}.
\]
By Chebyshev's inequality, for any $c$, we have
\[
P\left(\left|\frac{1}{\left|N_{n}(h_{k})\right|}\sum_{N_{n}(h_{k})}\left\Vert \Gamma_{i}^{\varepsilon}\right\Vert _{2}-E\left\Vert \Gamma_{i}^{\varepsilon}\right\Vert _{2}\right|\ge c\right)\le\frac{V\left(\left\Vert \Gamma_{i}^{\varepsilon}\right\Vert _{2}\right)}{\left|N_{n}(h_{k})\right|c^{2}}\rightarrow0,
\]
and thus,
\begin{equation}
\left|O\left(\lambda_{n}^{-\frac{d}{4}}\right)\right|\frac{1}{\left|N_{n}(h_{k})\right|}\sum_{N_{n}(h_{k})}\left\Vert \Gamma_{i}^{\varepsilon}\right\Vert _{2}=\left|O\left(\lambda_{n}^{-\frac{d}{4}}\right)\right|\left(E\left\Vert \Gamma_{i}^{\varepsilon}\right\Vert _{2}+o_{p}(1)\right)=O_{p}\left(\lambda_{n}^{-\frac{d}{4}}\right),\label{eq:summand2}
\end{equation}
because $E\left\Vert \Gamma_{i}^{\varepsilon}\right\Vert _{2}=O(1)$.

For the third summand of (\ref{eq:gamma delta}), note that
\begin{align*}
&\phantom{\le}
\frac{1}{\left|N_{n}(h_{k})\right|}\left\Vert \sum_{N_{n}(h_{k})}\left(\Gamma_{i}^{s}+\triangle_{i,j}^{s}\right)\triangle_{i,j}^{\varepsilon}\right\Vert _{2} \\
&\le\frac{1}{\left|N_{n}(h_{k})\right|}\sum_{N_{n}(h_{k})}\left\Vert \Gamma_{i}^{s}+\triangle_{i,j}^{s}\right\Vert _{2}\left\Vert \triangle_{i,j}^{\varepsilon}\right\Vert _{2}
\\
&\le C\frac{1}{\left|N_{n}(h_{k})\right|}\sum_{N_{n}(h_{k})}\left\Vert \triangle_{i,j}^{\varepsilon}\right\Vert _{2}.
\end{align*}
Because $(a-b)^{2}\le\left|a^{2}-b^{2}\right|$ generally, we have
\[
\left\Vert \triangle_{i,j}^{\varepsilon}\right\Vert _{2}=\left|\left(\varepsilon(s_{i})-\varepsilon(s_{i}+h_{k})\right)-\left(\varepsilon(s_{i})-\varepsilon(s_{j})\right)\right|\le\sqrt{\left|\left(\varepsilon(s_{i})-\varepsilon(s_{i}+h_{k})\right)^{2}-\left(\varepsilon(s_{i})-\varepsilon(s_{j})\right)^{2}\right|},
\]
and by Markov's inequality we have that, for any $\epsilon>0$, 
\begin{align*}
&\phantom{\le}
P\left(\lambda_{n}^{\frac{d}{4}}\sqrt{\left|\left(\varepsilon(s_{i})-\varepsilon(s_{i}+h_{k})\right)^{2}-\left(\varepsilon(s_{i})-\varepsilon(s_{j})\right)^{2}\right|}>\epsilon\right) \\
&
\le\frac{1}{\epsilon^{2}}\lambda_{n}^{\frac{d}{2}}E\left|\left(\varepsilon(s_{i})-\varepsilon(s_{i}+h_{k})\right)^{2}-\left(\varepsilon(s_{i})-\varepsilon(s_{j})\right)^{2}\right|\rightarrow0,
\end{align*}
meaning that $\sqrt{\left|\left(\varepsilon(s_{i})-\varepsilon(s_{i}+h_{k})\right)^{2}-\left(\varepsilon(s_{i})-\varepsilon(s_{j})\right)^{2}\right|}=o_{p}\left(\lambda_{n}^{-\frac{d}{4}}\right)$.
Consequently,
\begin{equation}
C\frac{1}{\left|N_{n}(h_{k})\right|}\sum_{N_{n}(h_{k})}\left\Vert \triangle_{i,j}^{\varepsilon}\right\Vert _{2}=o_{p}\left(\lambda_{n}^{-\frac{d}{4}}\right).\label{eq:summand3}
\end{equation}

Combining (\ref{eq:summand1}), (\ref{eq:summand2}), and (\ref{eq:summand3}),
we establish that (\ref{eq:gamma delta}) is
\[
n^{1/2}o_{p}\left(\lambda_{n}^{-\frac{d}{4}}\right)\left(o_{p}(\lambda_{n}^{-\frac{d}{4}})+O_{p}\left(\lambda_{n}^{-\frac{d}{4}}\right)+o_{p}\left(\lambda_{n}^{-\frac{d}{4}}\right)\right)=n^{1/2}o_{p}\left(\lambda_{n}^{-\frac{d}{2}}\right)
,\]
since $\left\Vert \hat{\rho}-\rho\right\Vert _{2}=o_{p}\left(\lambda_{n}^{-\frac{d}{4}}\right)$.

Consequently, combining with (\ref{eq:rhos}), we obtain
\[
\left|g_{1n}-g_{n}\right|=n^{1/2}o_{p}\left(\lambda_{n}^{-\frac{d}{2}}\right)=o_{p}(1).
\]

For the remainder of the proof, use the more economical notation
\begin{align*}
\mathcal{E}_{i}(k):=(\varepsilon(x_{i})-\varepsilon(x_{i}+h_{k}))^{2}-\gamma(h_{k};\theta_{0}),\quad i=1,...,n,
\end{align*}
and 
\useshortskip
\begin{align*}
\mathcal{E}_{(i,j)}(k):=(\varepsilon(x_{i})-\varepsilon(x_{j}))^{2}-\gamma(h_{k};\theta_{0}),\quad \forall\ (i,j)\in N_{n}(h_{k}),\quad k=1,...,K.
\end{align*}
Note that
\begin{align*}
E\left|g_{1n}-g_{2n}\right| & \le n^{1/2}\sum_{k=1}^{K}a_{k}E\left|\frac{1}{\left|N_{n}(h_{k})\right|}\sum_{(i,j)\in N_{n}(h_{k})}\mathcal{E}_{(i,j)}(k)-\frac{1}{n}\sum_{i=1}^{n}\mathcal{E}_{i}(k)\right|\\
 & =n^{1/2}\sum_{k=1}^{K}a_{k}E\left|\frac{1}{\left|N_{n}(h_{k})\right|}\sum_{(i,j)\in N_{r,n}(h_{k})}\mathcal{E}_{(i,j)}(k)
 \right.
 \\
 &\phantom{=}
 \quad
 +
 \quad
 \left.
 \frac{1}{\left|N_{n}(h_{k})\right|}\sum_{(i,j)\in N_{u,n}(h_{k})}\mathcal{E}_{(i,j)}(k)-\frac{1}{n}\sum_{i=1}^{n}\mathcal{E}_{i}(k)\right|\\
 & =n^{1/2}\sum_{k=1}^{K}a_{k}E\left|\frac{1}{\left|N_{n}(h_{k})\right|}\sum_{(i,j)\in N_{r,n}(h_{k})}\mathcal{E}_{(i,j)}(k)+\frac{1}{\left|N_{n}(h_{k})\right|}\sum_{(i,j)\in N_{u,n}(h_{k})}\left(\mathcal{E}_{(i,j)}(k)-\mathcal{E}_{i}(k)\right)\right.\\
 &\phantom{=} \quad
 \left. + \quad
 \frac{1}{\left|N_{n}(h_{k})\right|}\sum_{(i,j)\in N_{u,n}(h_{k})}\mathcal{E}_{i}(k)-\frac{1}{n}\sum_{i=1}^{n}\mathcal{E}_{i}(k)\right|\\
 & =n^{1/2}\sum_{k=1}^{K}a_{k}E\left|\frac{1}{\left|N_{n}(h_{k})\right|}\sum_{(i,j)\in N_{r,n}(h_{k})}\mathcal{E}_{(i,j)}(k)+\frac{1}{\left|N_{n}(h_{k})\right|}\sum_{(i,j)\in N_{u,n}(h_{k})}\left(\mathcal{E}_{(i,j)}(k)-\mathcal{E}_{i}(k)\right)\right.\\
 &\phantom{=} \quad
 \left. + \quad
 \left(\frac{1}{\left|N_{n}(h_{k})\right|}-\frac{1}{n}\right)\sum_{(i,j)\in N_{u,n}(h_{k})}\mathcal{E}_{i}(k)-\frac{1}{n}\sum_{i\in N'(h_{k})}\mathcal{E}_{i}(k)\right|\\
 & \le n^{1/2}\sum_{k=1}^{K}a_{k}\left(\frac{1}{\left|N_{n}(h_{k})\right|}E\left|\sum_{(i,j)\in N_{r,n}(h_{k})}\mathcal{E}_{(i,j)}(k)\right|+E\left|\mathcal{E}_{(i,j)}(k)-\mathcal{E}_{i}(k)\right|\right.\\
 &\phantom{=} \quad 
 \left. + \quad E\left|\left(\frac{1}{\left|N_{n}(h_{k})\right|}-\frac{1}{n}\right)\sum_{(i,j)\in N_{u,n}(h_{k})}\mathcal{E}_{i}(k)\right|+E\left|\frac{1}{n}\sum_{i\in N'(h_{k})}\mathcal{E}_{i}(k)\right|\right).
\end{align*}
We consider the summands one by one.

For the first summand,
\begin{align*}
\frac{1}{\left|N_{n}(h_{k})\right|}E\left|\sum_{(i,j)\in N_{r,n}(h_{k})}\mathcal{E}_{(i,j)}(k)\right| & \le\frac{1}{\left|N_{n}(h_{k})\right|}E\left|\sum_{(i,j)\in N_{r,n}(h_{k})}\mathcal{E}_{i}(k)\right|
\\
 &\phantom{=}
 \quad
 +
 \quad
 \frac{1}{\left|N_{n}(h_{k})\right|}E\left|\sum_{(i,j)\in N_{r,n}(h_{k})}\mathcal{E}_{(i,j)}(k)-\mathcal{E}_{i}(k)\right|\\
 & \le\frac{1}{\left|N_{n}(h_{k})\right|}\left(E\left(\sum_{(i,j)\in N_{r,n}(h_{k})}\mathcal{E}_{i}(k)\right)^{2}\right)^{\frac{1}{2}}
 \\
 &\phantom{=}
 \quad
 +
 \quad
 \frac{1}{\left|N_{n}(h_{k})\right|}\sum_{(i,j)\in N_{r,n}(h_{k})}E\left|\mathcal{E}_{(i,j)}(k)-\mathcal{E}_{i}(k)\right|\\
 & \le\frac{1}{\left|N_{n}(h_{k})\right|}O\left(\lambda_{n}^{-\frac{d}{2}}\left|N_{r,n}(h_{k})\right|\right)+\frac{\left|N_{r,n}(h_{k})\right|}{\left|N_{n}(h_{k})\right|}o\left(\lambda_{n}^{-\frac{d}{2}}\right)\\
 & =O(1)\frac{1}{n}o(\lambda_{n}^{d})O\left(\lambda_{n}^{-\frac{d}{2}}\right)+\frac{O(1)}{n}o(\lambda_{n}^{d})o\left(\lambda_{n}^{-\frac{d}{2}}\right)=\frac{1}{n}o\left(\lambda_{n}^{\frac{d}{2}}\right),
\end{align*}
where we used the fact that $1/\left|N_{n}(h_{k})\right|=1/(n(1+o(1))=1/n\cdot O(1)$.

For the second summand, we have by assumption that $E\left|\mathcal{E}_{(i,j)}(k)-\mathcal{E}_{i}(k)\right|=o\left(\lambda_{n}^{-\frac{d}{2}}\right)$.

For the third summand,
\begin{align*}
\left|\frac{1}{\left|N_{n}(h_{k})\right|}-\frac{1}{n}\right|\cdot E\left|\sum_{(i,j)\in N_{u,n}(h_{k})}\mathcal{E}_{i}(k)\right| & =\frac{O(1)}{n^{2}}\left|\left|N_{n}(h_{k})\right|-n\right|\cdot\left(E\left(\sum_{(i,j)\in N_{u,n}(h_{k})}\mathcal{E}_{i}(k)\right)^{2}\right)^{\frac{1}{2}}\\
 & =\frac{o(1)}{n}\cdot O\left(\lambda_{n}^{-\frac{d}{2}}\left|N_{u,n}(h_{k})\right|\right)\\ & \le \frac{o(1)}{n}\cdot \left|N_{n}(h_{k})\right|\cdot O\left(\lambda_{n}^{-\frac{d}{2}}\right) \\ & =  \frac{o(1)}{n} \cdot  n(1+o(1))\cdot O\left(\lambda_{n}^{-\frac{d}{2}}\right) =o\left(\lambda_{n}^{-\frac{d}{2}}\right).
\end{align*}
For the fourth summand,
\begin{align*}
\frac{1}{n}E\left|\sum_{i\in N'(h_{k})}\mathcal{E}_{i}(k)\right| & \le\frac{1}{n}\left(E\left(\sum_{i\in N'(h_{k})}\mathcal{E}_{i}(k)\right)^{2}\right)^{\frac{1}{2}}\\
 & =\frac{1}{n}O\left(\left|N'(h_{k})\right|\lambda_{n}^{-\frac{d}{2}}\right)=\frac{1}{n}o\left(\lambda_{n}^{\frac{d}{2}}\right).
\end{align*}

Considered altogether, we have
\begin{align*}
E\left|g_{1n}-g_{2n}\right| & =n^{1/2}\left(\frac{1}{n}o\left(\lambda_{n}^{\frac{d}{2}}\right)+o\left(\lambda_{n}^{-\frac{d}{2}}\right)+o\left(\lambda_{n}^{-\frac{d}{2}}\right)+\frac{1}{n}o\left(\lambda_{n}^{\frac{d}{2}}\right)\right)\\
 & =n^{-1/2}o\left(\lambda_{n}^{\frac{d}{2}}\right)+n^{1/2}o\left(\lambda_{n}^{-\frac{d}{2}}\right).
\end{align*}

Consequently, under pure-increasing domain asymptotics, $E\left|g_{1n}-g_{2n}\right|=o(1)$.
Then it suffices to show that $g_{2n}\rightarrow N(0,a^{T}\Sigma_{g}(\theta_{0})a)$,
which follows from an application of Lemma \ref{lem:b1}.

To obtain (ii), for the mixed-increasing domain, let $\lambda_{n}^{\frac{d}{2}}a^{T}g_{n}(\phi_{0})\overset{d}{\rightarrow}N\left(0,a^{T}\Sigma_{g}(\phi_{0})a\right)$
as $n\rightarrow\infty$ for any $a\in\mathbb{R}^{k}$. Let $g_{0n}=\lambda_{n}^{\frac{d}{2}}a^{T}g_{n}(\phi_{0})$,
and likewise redefine 
\[
g_{1n}=\lambda_{n}^{\frac{d}{2}}\sum_{k=1}^{K}a_{k}\left(\frac{1}{\left|N_{n}(h_{k})\right|}\sum_{N_{n}(h_{k})}\left(\varepsilon(x_{i})-\varepsilon(x_{j})\right)^{2}-\gamma(h_{k};\phi_{0})\right),
\]
and 
\[
g_{2n}=\lambda_{n}^{\frac{d}{2}}\sum_{k=1}^{K}a_{k}\left(\frac{1}{n}\sum_{i=1}^{n}\left(\varepsilon(x_{i})-\varepsilon(x_{i}+h_{k})\right)^{2}-\gamma(h_{k};\phi_{0})\right).
\]
It is then immediate from the above argument that 
\[
\left|g_{1n}-g_{n}\right|=\lambda_{n}^{\frac{d}{2}}o_{p}\left(\lambda_{n}^{-\frac{d}{2}}\right)=o_{p}(1),
\]
and
\[
E\left|g_{1n}-g_{2n}\right|=\lambda_{n}^{\frac{d}{2}}\left(\frac{1}{n}o\left(\lambda_{n}^{\frac{d}{2}}\right)+o\left(\lambda_{n}^{-\frac{d}{2}}\right)\right)=o\left(\lambda_{n}^{d}/n\right)+o\left(1\right)=o\left(1\right).
\]
\end{Proof}

The proof for $g_{n}$ built, in addition, with covariogram $\frac{1}{|N_{n}(d)|}\sum_{(i,j)\in N_{n}(d)}(\hat{R}(x_{i})-Y(x_{j}))^{2}$
is carried out analogously.

With the asymptotic distribution of the statistic $g_{n}$ in hand,
we can call on Theorem 3.2 of Cressie and Lahiri (2002) and obtain
Corollary 1 as an immediate corollary.

\section{Two-step bootstrap implementation with estimated linear mean} \label{app:bootstrap}

If $m(x)$ is modeled as $s(x)^{T}\rho$ instead of as a constant
$m$, then the level of the imputed mean $\hat{m}(x)$ may affect
the estimated regression coefficient $\hat{\beta}$ and must be accounted
for in the two-step bootstrap procedure.

Let $\vartheta=(\rho,\theta)$. The procedure may be extended as:
\begin{itemize}
\item Draw $\hat{\vartheta}^{(j)}\sim N\left(\hat{\vartheta}_{\mathrm{mle}},\widehat{V\left(\hat{\vartheta}_{\mathrm{mle}}\right)}\right)$,
i.e., from its asymptotic distribution (using only $\mathbf{R}^{*}$
as data)
\item Compute
\useshortskip
\begin{align*}
\hat{\mathbf{R}}^{(j)}=\hat{\mathbf{R}}(\hat{\theta}^{(j)})=\mathbf{m}_{\hat{\rho}^{(j)}}+\bar{\mathbf{K}}_{\hat{\theta}^{(j)}}^{T}\mathbf{K}_{\hat{\theta}^{(j)}}^{*-1}\left(\mathbf{R}^{*}-\mathbf{m}_{\hat{\rho}^{(j)}}^{*}\right),
\end{align*}
where $\mathbf{m}_{\hat{\rho}^{(j)}}=\left(s(x_{1})^{T}\hat{\rho}^{(j)},...,s(x_{N})^{T}\hat{\rho}^{(j)}\right)$
and likewise for $\mathbf{m}_{\hat{\rho}^{(j)}}^{*}$
\item Draw new data set $\mathscr{D}^{(j)}$ with replacement from $\left(\mathbf{Y},\hat{\mathbf{R}}^{(j)}\right)$
\item Calculate $\hat{\beta}^{(j)}$, the regression coefficient for the
data set $\mathscr{D}^{(j)}$
\end{itemize}
For moderate or large $\mathrm{dim}(\rho)$, maximum likelihood estimation
of $\vartheta$ will produce a poor estimate of $\theta$, and REML
methods are usually preconized for estimating $\theta$. We thus suggest
to estimate $\vartheta$ and sample $\hat{\vartheta}$ according to a pseudo likelihood. Specifically, $\hat{\rho}^{(j)}$ is drawn from
a normal distribution with mean $\hat{\rho}$ and variance $\frac{1}{n}\left(s(x^{*})^{T}s(x^{*})\right)^{-1}s(x^{*})^{T}\mathbf{K}^{*}s(x^{*})\left(s(x^{*})^{T}s(x^{*})\right)^{-1}$
if $\hat{\rho}$ is estimated by OLS, and with variance $\frac{1}{n}\left(s(x^{*})^{T}\mathbf{K}^{*}s(x^{*})\right)^{-1}$
is $\hat{\rho}$ is estimated by GLS.  The covariance parameter $\theta$ is estimated by REML and  $\hat{\theta}^{(j)}$ is sampled from the corresponding asymptotic distribution \citep{gelfand2010handbook}. This is how
we implement the sampling in the companion R package SpReg, which is freely available. 
\end{document}